\documentclass[twocolumn]{aastex61} 
\usepackage{graphicx}				
\usepackage{amsmath}					
\usepackage{amssymb}
\usepackage{graphbox}
\usepackage{verbatim}

\begin{document}

\title{A search for optical AGN variability in $35,000$ low-mass galaxies with the Palomar Transient Factory}
\author{Vivienne F. Baldassare}
\altaffiliation{Einstein Fellow}
\affiliation{Yale University \\
Department of Astronomy \\
52 Hillhouse Avenue \\
New Haven, CT 06511, USA}

\author{Marla Geha}
\affiliation{Yale University \\
Department of Astronomy \\
52 Hillhouse Avenue \\
New Haven, CT 06511, USA}

\author{Jenny Greene}
\affiliation{Princeton University \\
Department of Astrophysical Sciences\\
4 Ivy Lane\\
Princeton University, Princeton, NJ 08544 }

\correspondingauthor{Vivienne F. Baldassare} \email{vivienne.baldassare@yale.edu}

\received{13 October 2019}
\submitjournal{ApJ}

\begin{abstract}

We present an analysis of the long-term optical variability for $\sim50,000$ nearby ($z<0.055$) galaxies from the NASA-Sloan Atlas, $35,000$ of which are low-mass ($M_{\ast}<10^{10}~M_{\odot}$). We use difference imaging of Palomar Transient Factory (PTF) R-band observations to construct light curves with typical baselines of several years. We then search for subtle variations in the nuclear light output. We determine whether detected variability is AGN-like by assessing the fit quality to a damped random walk model. We identify 424 variability-selected AGN, including 244 with stellar masses between $10^{7}$ and $10^{10}~M_{\odot}$. 75\% of low-mass galaxies with AGN-like variability have narrow emission lines dominated by star formation. After controlling for nucleus magnitude, the fraction of variable AGN is constant down to $M_{\ast}=10^{9}~M_{\odot}$, suggesting no drastic decline in the BH occupation fraction down to this  stellar mass regime. Combining our NASA-Sloan Atlas sample with samples of nearby galaxies with broad H$\alpha$ emission, we find no dependence of variability properties with black hole mass. However, we caution that the variable AGN fraction is strongly dependent on baseline. For baselines less than two years, the variable fraction for the full sample is 0.25\%, compared to 1.0\% for baselines longer than two years.   Finally, comparing Stripe 82 light curves (Baldassare et al. 2018) to PTF light curves, we find populations of changing-look AGN: 8 galaxies that are variable in Stripe 82, but quiescent in PTF, and 15 galaxies where the reverse is true. Our PTF work demonstrates the promise of long-term optical variability searches in low-mass galaxies for finding AGNs missed by other selection techniques.

\end{abstract}

\section{Introduction}

The last decade has seen a drastic increase in the number of publicly available repeat imaging surveys. This has enabled the detection and characterization of a wide range of transient and variable phenomena, including supernovae, tidal disruption events, and active galactic nuclei (AGNs). AGNs are observed to vary across the electromagnetic spectrum on time scales ranging from hours to years. Variability itself has long been a prolific tool for identifying AGN \citep{1997ARA&A..35..445U, 2003AJ....125....1G, 2007AJ....134.2236S, 2010ApJ...714.1194S, 2011ApJ...728...26M, 2014ApJ...782...37C, 2016ApJ...826...62H, 2017ApJ...834..111C}. 

The identification of AGNs via long-term optical variability is particularly interesting in the context of AGNs in low-mass galaxies ($M_{\ast}<10^{10}~M_{\odot}$). Massive black holes (BHs; $M_{\rm BH}\gtrsim10^{4}~M_{\odot}$) in low-mass galaxies are elusive; their relatively small sphere of influence makes them infeasible to find dynamically beyond $\sim5$ Mpc (see \citealt{2015ApJ...809..101D, 2017ApJ...836..237N, 2018ApJ...858..118N, 2019ApJ...872..104N} for examples of dynamical detections in low-mass galaxies within 5 Mpc). Thus, in general, searches for BHs in low-mass galaxies focus on signs of BH accretion rather than on dynamical signatures.

The first efforts to identify AGNs in low-mass galaxies in significant numbers used optical spectroscopic selection techniques. Using Sloan Digital Sky Survey (SDSS) data, \cite{2004ApJ...610..722G, 2007ApJ...670...92G} searched for broad H$\alpha$ emission indicative of low-mass black holes with $M_{\rm BH}\lesssim10^{6}~M_{\odot}$. \cite{Reines:2013fj} analyzed SDSS spectroscopy for 25,000 emission line galaxies with stellar masses less than the Large Magellanic Cloud (i.e., $M_{\ast}<3\times10^{9}~M_{\odot}$), and found 136 with optical emission line ratios indicative of AGN activity (based on the BPT diagram; \citealt{1981PASP...93....5B, 2003MNRAS.346.1055K, 2006MNRAS.372..961K}). 

While optical emission line ratios are a secure method of identifying AGNs in low-mass galaxies \citep{2016ApJ...829...57}, there is likely a population that is undetected due to selection effects that are increasingly relevant at low stellar masses. In particular, star formation dilution and low-metallicity effects are thought to compromise the detection of AGNs in low-mass galaxies. Star formation can dilute the contribution of a weak and/or low-mass AGN to the optical emission lines. This is especially true when considering the 3$''$ spectroscopic fiber of the SDSS, which sometimes encloses the entire spatial extent of a dwarf galaxy \citep{2015ApJ...811...26T, 2019arXiv190201401D}. Additionally, low galaxy metallicity has the effect of lowering the $\rm{[NII]}$-to-H$\alpha$ ratio and pushing objects to the left and out of the AGN regime on the classic BPT diagram \citep{2006MNRAS.371.1559G, 2019ApJ...870L...2C}.

In \cite{2018ApJ..868..152}, we used SDSS Stripe 82 data to search for low-level optical photometric variability characteristic of AGNs in a sample of $\sim28,000$ galaxies with stellar masses from $10^{7}-10^{12}~M_{\odot}$ from the NASA-Sloan Atlas. We found 135 galaxies with AGN-like optical photometric variability, as determined by the goodness-of-fit to a damped random walk model \citep{2009ApJ...698..895K, 2010ApJ...708..927K, 2011AJ....141...93B}. The variable AGN host galaxies ranged in stellar mass from $\sim10^{8}-10^{11}~M_{\odot}$. Interestingly, there was a difference in the optical spectroscopic properties of the high-mass ($M_{\ast}>10^{10}~M_{\odot}$) and low-mass ($M_{\ast}<10^{10}~M_{\odot}$) subsamples. Among the high-mass galaxies, almost 100\% had narrow-line ratios placing them in the AGN or composite regions of the BPT diagram. On the other hand, 50\% of the low-mass galaxies had narrow emission line ratios placing them in the star forming region of the BPT diagram, indicating that optical variability can identify AGN missed by other selection techniques. In the era of the Large Synoptic Survey Telescope (LSST; \citealt{2008SerAJ.176....1I}), which will image the entire visible sky every three nights, this represents an incredibly promising technique for finding substantial numbers of obscured or otherwise undetected AGN in low-mass galaxies. 

In this paper, we expand our search for variable AGN to the Palomar Transient Factory, which has covered almost the entire northern sky ($\sim30,000$ deg$^{2}$, compared to 275 deg$^{2}$ for SDSS Stripe 82). In Section 2, we describe our data set and samples of galaxies. In Section 3, we discuss the difference imaging and light curve construction techniques, and the selection of variable AGN. In Section 4, we present our variability results. In Section 5, we explore possible relations between BH mass and variability properties. Finally, in Section 6, we compare our sample of low-mass galaxies with AGN-like variability from PTF to those selected via other techniques.

\section{Data}

\subsection{Palomar Transient Factory}

The Palomar Transient Factory (PTF; \citealt{2009PASP..121.1334R, 2009PASP..121.1395L}) is a wide-field optical survey to study the transient and variable sky. PTF began operations in 2009, using the 48-inch Samuel Oschin Telescope at Palomar Observatory. PTF observed primarily in R-band with an exposure time of 60s and typical $5-\sigma$ limiting magnitude of $m_{R}=20.5$ mag. The pixel size of the CCD is $1''/\rm{pix}$. Observations were also made in g-band, but with significantly fewer epochs (the limiting g-band magnitude is $m_{g}=21$). The Intermediate Palomar Transient Survey (iPTF), which built upon the PTF, improved upon data reduction and source classification using an upgraded camera. The PTF/iPTF have observed almost the entire northern sky, with some parts of the sky being imaged several thousand times. Since the PTF/iPTF has pursued various experiments and observer-proposed projects, the cadence and baseline for any given region can differ widely from the rest of the survey. These data are publicly available, and the science addressed by the PTF/iPTF span a wide range of astrophysical phenomena including supernovae \citep{2012PASP..124.1175B, 2012ApJ...753...22B, 2016ApJ...830...13P}, cataclysmic variables, tidal disruption events \citep{2014ApJ...793...38A, 2019ApJ...872..198V}, and AGN \citep{2009PASP..121.1334R,2016ApJ...826...62H, 2017ApJ...835..144G}. 

\subsection{Galaxy samples}

We are interested in finding BHs in low-mass galaxies. In order to have accurate galaxy stellar mass estimates, we require the sample to have spectroscopic redshifts. We use the NASA-Sloan Atlas (NSA) as our parent galaxy sample. The NSA is a reprocessing of five-band SDSS DR8 photometry, combined with UV photometry from GALEX \citep{2011ApJS..193...29A, 2011AJ....142..153Y, 2007AJ....133..734B, 2011AJ....142...31B}. Most of the galaxies in the NSA have SDSS spectroscopy, though a small number have spectroscopic redshifts from e.g., the NASA Extragalactic Database. The NSA catalog provides a wealth of derived quantities, including spectroscopic redshifts, galaxy mass, radius, ellipticity, magnitude, and line flux measurements. We use NSA version 0, which extends out to z=0.055 (hereafter, referred to as NSAv0). Note that the ID numbers assigned to galaxies in the NSAv0 differ from those assigned in the NSA version 1, which extends to z=0.15. There are more than 145,000 galaxies in the NSAv0, with a median stellar mass of $4\times10^{9}~M_{\odot}$. 

Our Stripe 82 results showed that above $10^{10}~M_{\odot}$, almost all of the variable galaxies were also classified as AGN based on their optical emission line ratios. Thus, given limited resources, we focused our PTF analysis on low-mass galaxies. Our primary NSAv0 sample was constructed to include all galaxies with stellar masses $M_{\ast}\lesssim2\times10^{10}M_{\odot}$. There are \textcolor{black}{78872} galaxies below this mass cut in the NSA with some amount of coverage in the PTF. Additionally, we construct light curves for any NSA galaxy nearby enough to the target galaxy that they are contained within the field surrounding the target galaxy. 

After placing constraints on the minimum required number of data points (Section 3.1), our final NSAv0 analysis sample consists of \textcolor{black}{47125} galaxies spanning from $\sim10^{7}-10^{12}~M_{\odot}$, with a median stellar mass of $4.5\times10^{9}~M_{\odot}$ and a median redshift of $z=0.035$. This includes \textcolor{black}{35073} galaxies below $10^{10}~M_{\odot}$, almost 10 times as many low-mass galaxies as were in \cite{2018ApJ..868..152} within $z<0.055$.

In order to search for trends with BH mass, we also analyze data from the \cite{2007ApJ...670...92G} catalog of low-mass ($M_{\rm BH} \lesssim 10^{6}~M_{\odot}$) broad-line AGN, and the \cite{2015ApJ...813...82R} compilation of broad-line AGNs in the NSAv0. Finally, we include any variability-selected AGN identified in our Stripe 82 study \citep{2018ApJ..868..152}; this includes high and low-mass galaxies out to z=0.15.

\section{Data Analysis}

\begin{figure}
\includegraphics[width=0.5\textwidth]{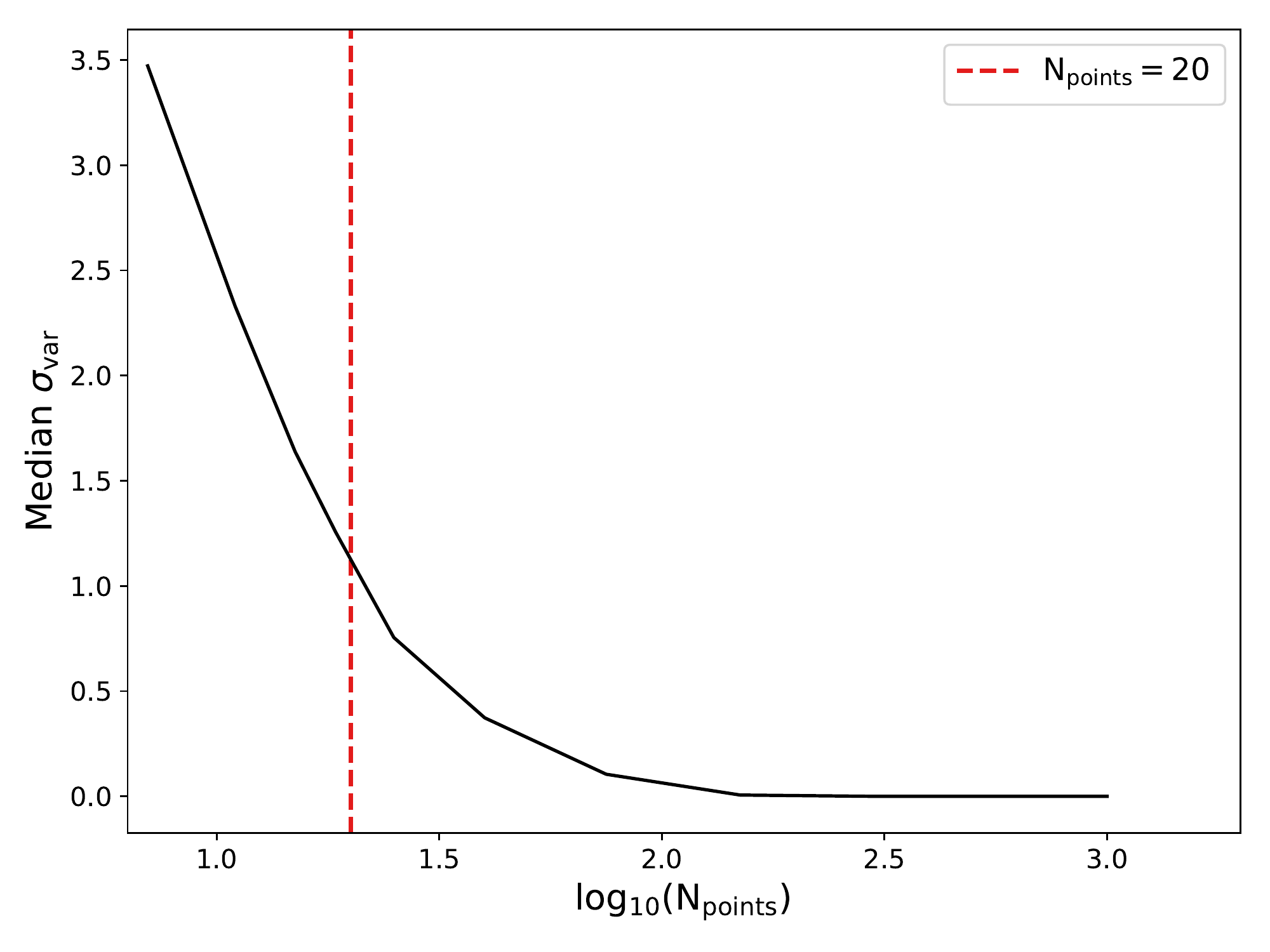}
\caption{Median $\sigma_{\rm var}$ versus number of data points in a light curve. There is a clear leveling off of the median variability significance beyond $\sim20$ data points. With N$\lesssim$20, the median light curve $\sigma_{\rm var}$ is greater than $1-\sigma$; we do not use light curves with less than 20 data points in our analysis. }
\label{sig_v_N} 
\end{figure}

\subsection{Difference imaging and light curve construction}

Our goal is to detect low level variability (less than a tenth of a magnitude) of point sources super-posed on top of extended host galaxies. We use difference imaging to subtract off the host galaxy light before measuring any light from the galaxy nucleus. Our analysis is similar to \cite{2018ApJ..868..152}, and summarized here. 

For each target galaxy we download $300''\times300''$ PTF R-band images (roughly $300\times300$ pixels), centered on the target. We use only PTF images that have seeing better than 3$''$. We use the R-band as opposed to g-band, since the sky coverage, baselines, and number of observations  are substantially better in the R-band. After retrieving all R-band images for a given target from the PTF database, we proceed with difference imaging. 

We use the software Difference Imaging and Analysis Pipeline 2 (DIAPL2), which is a modified version of the Difference Imaging Analysis software \citep{2000AcA....50..421W}. Both are based on the difference imaging analysis introduced by \cite{1998ApJ...503..325A} and \cite{2000A&AS..144..363A}. The first step is to construct a template image by combining the best frames (i.e., those with the best seeing and lowest background). Then, for each individual exposure, the template image is convolved with a best fit kernel to match the seeing of that exposure. The kernel is a sum of 2D Gaussians of different widths. The template background is also matched to that of the exposure. Finally, the convolved template is subtracted from the exposure to create a difference image. 

To construct light curves for each galaxy, we carry out aperture photometry on the template and difference images. Thus, the flux value for each data point is the template value plus the difference image value. We use an aperture of 3$''$ centered on the galaxy nucleus as given in the NSAv0. This aperture was chosen to match the seeing of the lowest quality frames we use. 

In Figure~\ref{sig_v_N}, we show the median $\sigma_{\rm var}$ (or the significance that a galaxy is variable) versus the number of data points in a light curve. Based on this figure, we see that for targets with less than $\sim20$ data points, the median measured $\sigma_{\rm var}$ is $>1-\sigma$ (i.e., they are more likely to be flagged as variable). This is because, with few data points, there are fewer high quality images from which to construct a template, and an overall poorer image subtraction result. We choose to exclude light curves with less than 20 points from all further analysis. This reduces our sample from 78872 objects to 47125. 

\begin{figure}
\includegraphics[width=0.45\textwidth]{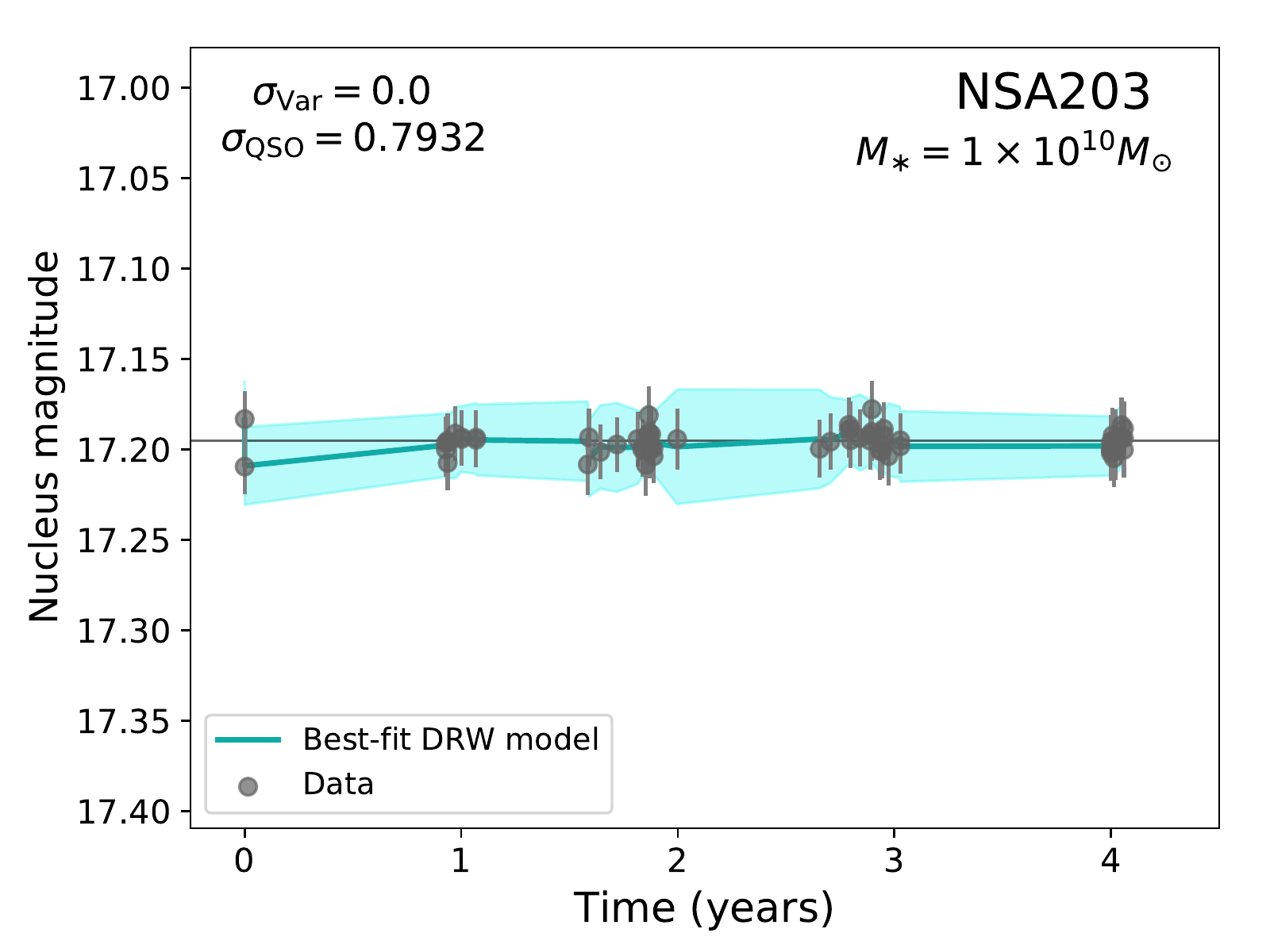}
\includegraphics[width=0.45\textwidth]{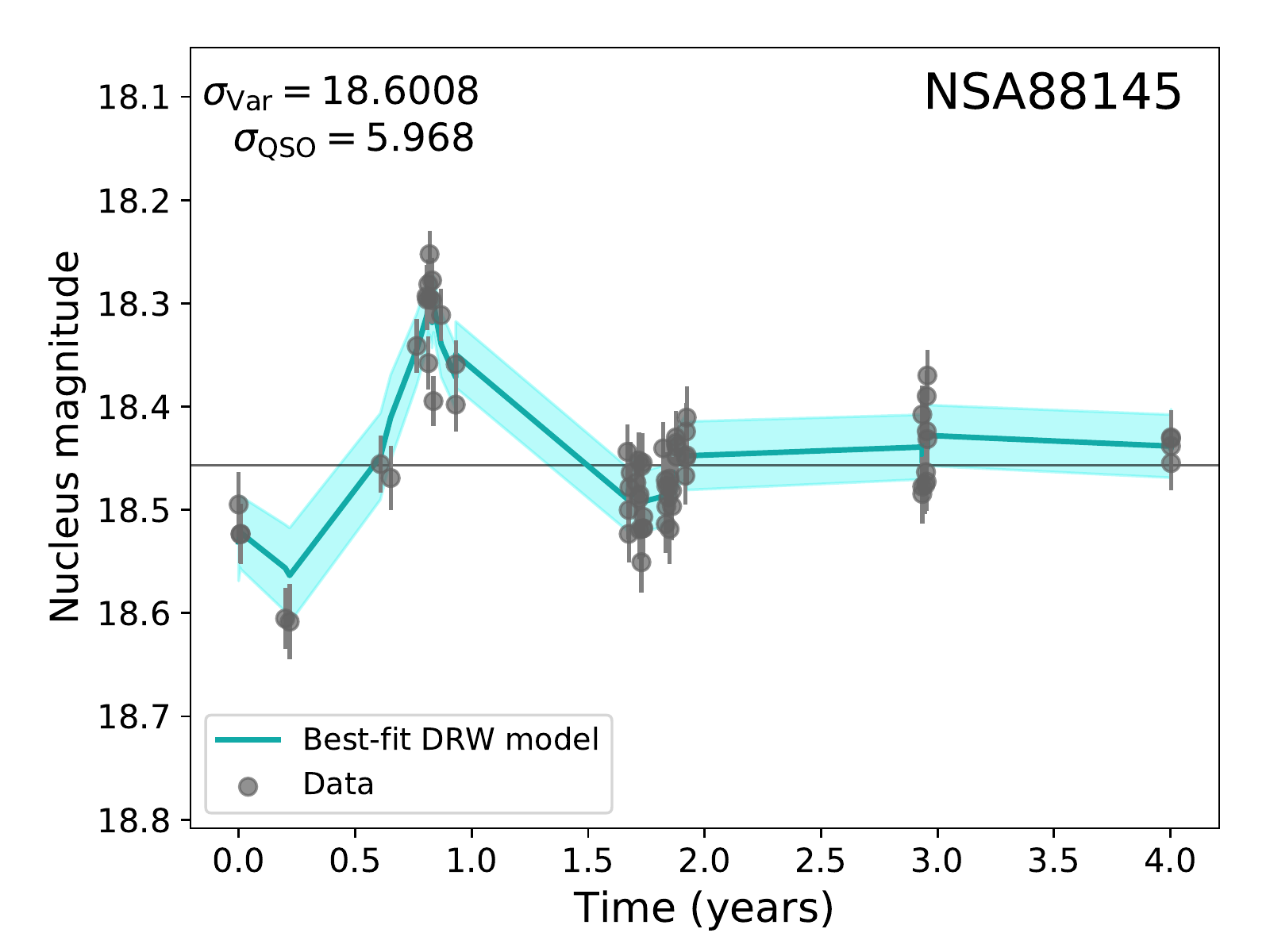}
\caption{Representative light curves for one non-variable galaxy (NSA 203) and one variable galaxy (NSA88145). The measured nucleus magnitude values are shown as gray circles. The best fit damped random walk model is given by the blue line, while the blue shaded region represents the model uncertainties. Both light curves span $\sim4$ years and contain $\sim70$ data points, which are values typical of the PTF sample.  }
\label{nonvar}
\end{figure}

Light curve baselines for the final sample range from 3 days to 2156 days, with a median baseline of 1474 days.  The median number of data points is 65, and the maximum number of points in a light curve is 1625. Representative light curves for variable and non-variable galaxies are shown in Figure~\ref{nonvar}. 

\subsection{Detection of AGN-like variability}

We select galaxies with AGN-like variability based on the goodness of fit of the galaxy's light curve to a damped random walk (DRW) model. The DRW is generally a good empirical descriptor of AGN variability \citep{2009ApJ...698..895K, 2010ApJ...721.1014M, 2010ApJ...708..927K, 2011AJ....141...93B}. In order to select targets which are both (i) variable and (ii) have AGN-like variability, we use the software QSO\_fit \citep{2011AJ....141...93B}. This software takes the measured magnitude values and dates as input and returns a best-fit DRW model, along with the quantities $\sigma_{\rm var}$, $\sigma_{\rm QSO}$, and $\sigma_{\rm notQSO}$. $\sigma_{\rm var}$ is the significance that the object is variable; $\sigma_{\rm QSO}$ is the significance that the $\chi^{2}$ for the damped random walk model is better than the expected $\chi^{2}$ for non AGN-like variability; $\sigma_{\rm notQSO}$ is the significance that the source variability is better described by random variability. 

We select objects as candidate variable AGN if they have $\sigma_{\rm var}>2$, $\sigma_{\rm QSO}>2$, and $\sigma_{\rm QSO}\gtrsim \sigma_{\rm notQSO} $. All light curves and difference images are then inspected by eye to remove spurious variability detections due to e.g., poor difference imaging or bad image frames. In all, we find \textcolor{black}{424} galaxies with AGN-like variability out of \textcolor{black}{47125} total analyzed objects. 

\begin{figure*}
\centering
\includegraphics[width=0.7\textwidth]{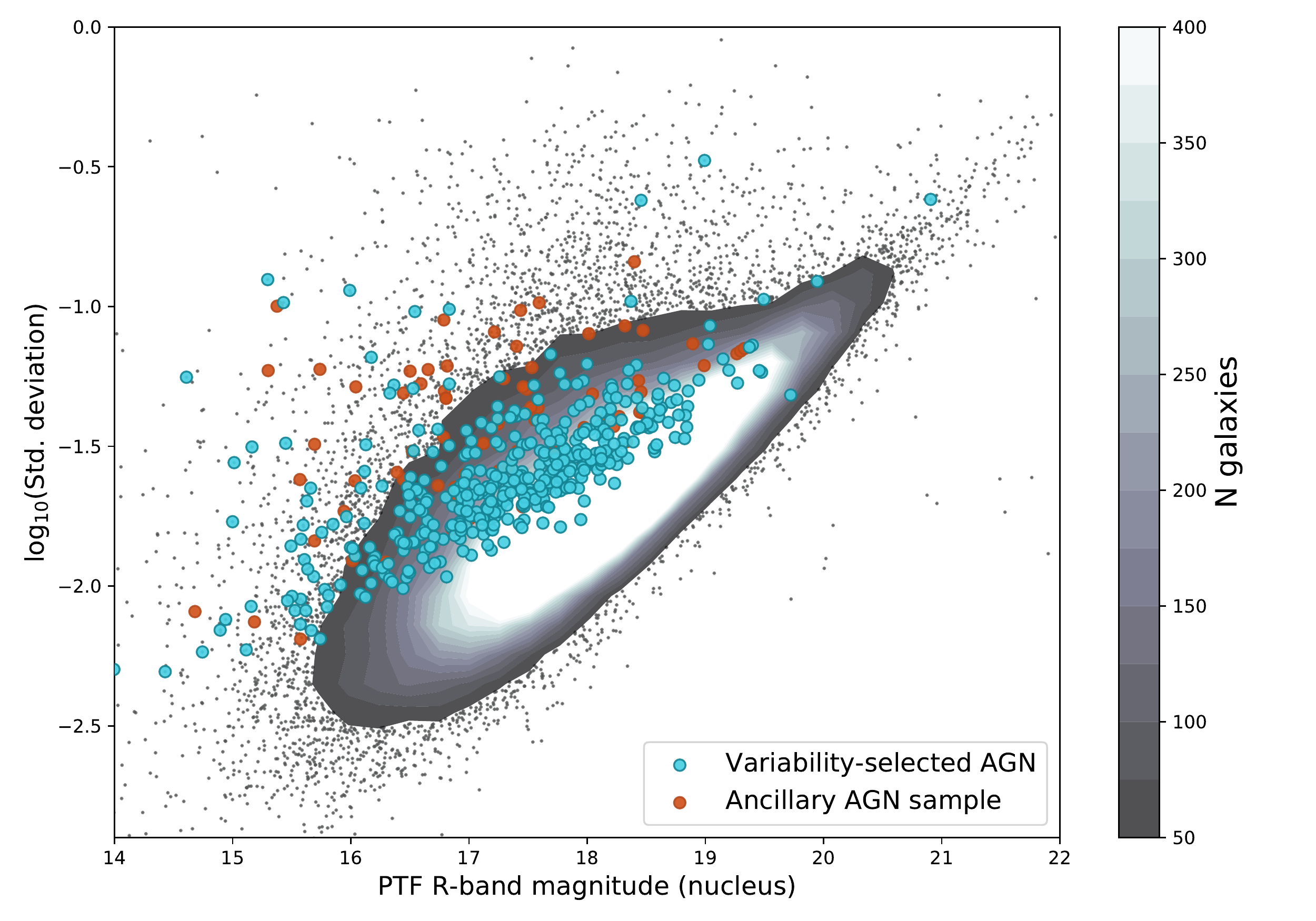}
\caption{Light curve standard deviation versus PTF R-band nuclear magnitude. The shaded gray contours show the full population of galaxies from the NSAv0. The blue circles show galaxies we identify as having AGN-like variability. The orange points show the variable AGN from our ancillary samples (described in Section 2.2). The typical scatter in a light curve increases towards fainter galaxy nuclei.}
\label{std_v_mag}
\end{figure*}

\subsection{Optical spectroscopic analysis}

We re-analyze SDSS optical spectroscopy for all galaxies found to have AGN-like variability to search for narrow and broad emission line signatures of AGN activity. Following the same approach as \cite{2018ApJ..868..152} (see also \citealt{Reines:2013fj, 2015ApJ...809L..14B, 2016ApJ...829...57}), we model the $\rm{H}\beta$, $[\rm{OIII}] \lambda 5007$, $\rm{H}\alpha$, and $[\rm{NII}] \lambda 6548, 6583$ emission lines. First, we create a narrow emission line model using the [S II] $\lambda\lambda$ 6713 and 6731 lines. These are forbidden transitions and thus not produced in the denser broad line region. We then use the width of the [S II] lines to fit the narrow H$\alpha$ emission and the [NII] $\lambda\lambda$6548,6684 lines simultaneously. The width of narrow H$\alpha$ is allowed to increase by 25\% relative to the [SII] width, and the relative amplitudes of [NII] $\lambda\lambda$6548,6684 are fixed to laboratory values. We next add a broader Gaussian component representing broad H$\alpha$ emission to the model. The component is kept as part of the model if the $\chi^{2}$ value of the fit improves by 20\%. If the spectrum is better fit with a broad H$\alpha$ component, we also test a model with two Gaussian components to represent broad H$\alpha$, again keeping the additional Gaussian if the $\chi^{2}$ value of the fit improves by at least 20\%.  We also fit H$\beta$ and [OIII] $\lambda$5007. H$\beta$ can also be fit with an additional broad component. In all of the line fits, the continuum is represented by a line fit across the relevant spectral region. 

If broad H$\alpha$ is present, we estimate the BH mass using the full width half maximum (FWHM) and luminosity of the broad H$\alpha$ component \citep{2005ApJ...630..122G}. Assuming the gas in the broad line region is virialized, we can estimate the BH mass with the distance to the broad line region, and the velocity of the broad line region gas. The FWHM of broad H$\alpha$ gives an estimate of the velocity of the gas in the broad line region, and the luminosity of broad H$\alpha$ has been found to be correlated with the distance from the BH to the broad line region \citep{2005ApJ...630..122G, 2009ApJ...705..199B, 2013ApJ...767..149B}. BH masses estimated via this technique have typical uncertainties of $\sim0.3$ dex.

\section{Variability Results}

\begin{figure*}
\centering
 \begin{minipage}{.26\textwidth}
        \centering
        \includegraphics[width=\textwidth]{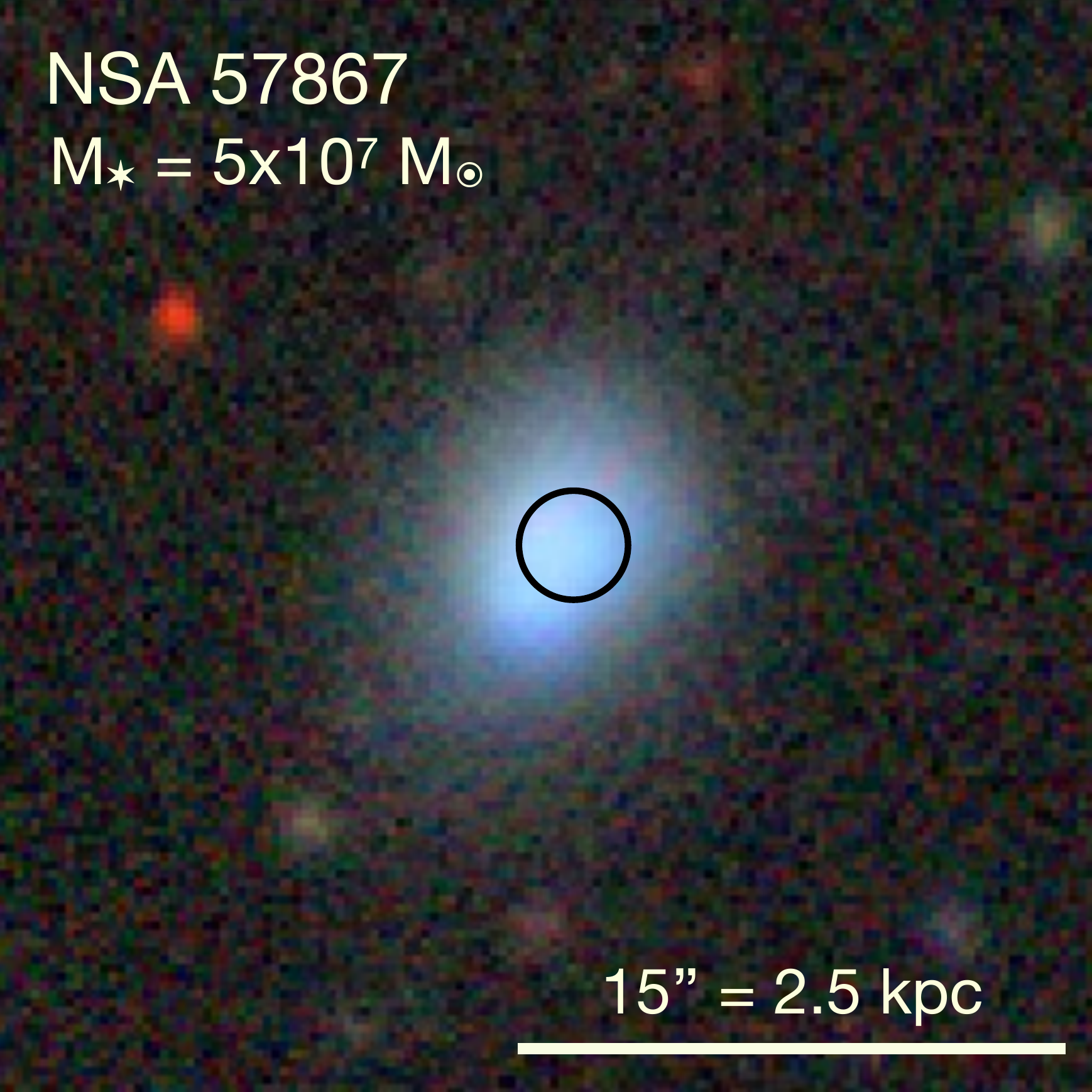}
    \end{minipage}%
    \begin{minipage}{0.55\textwidth}
        \centering
        \includegraphics[width=\textwidth]{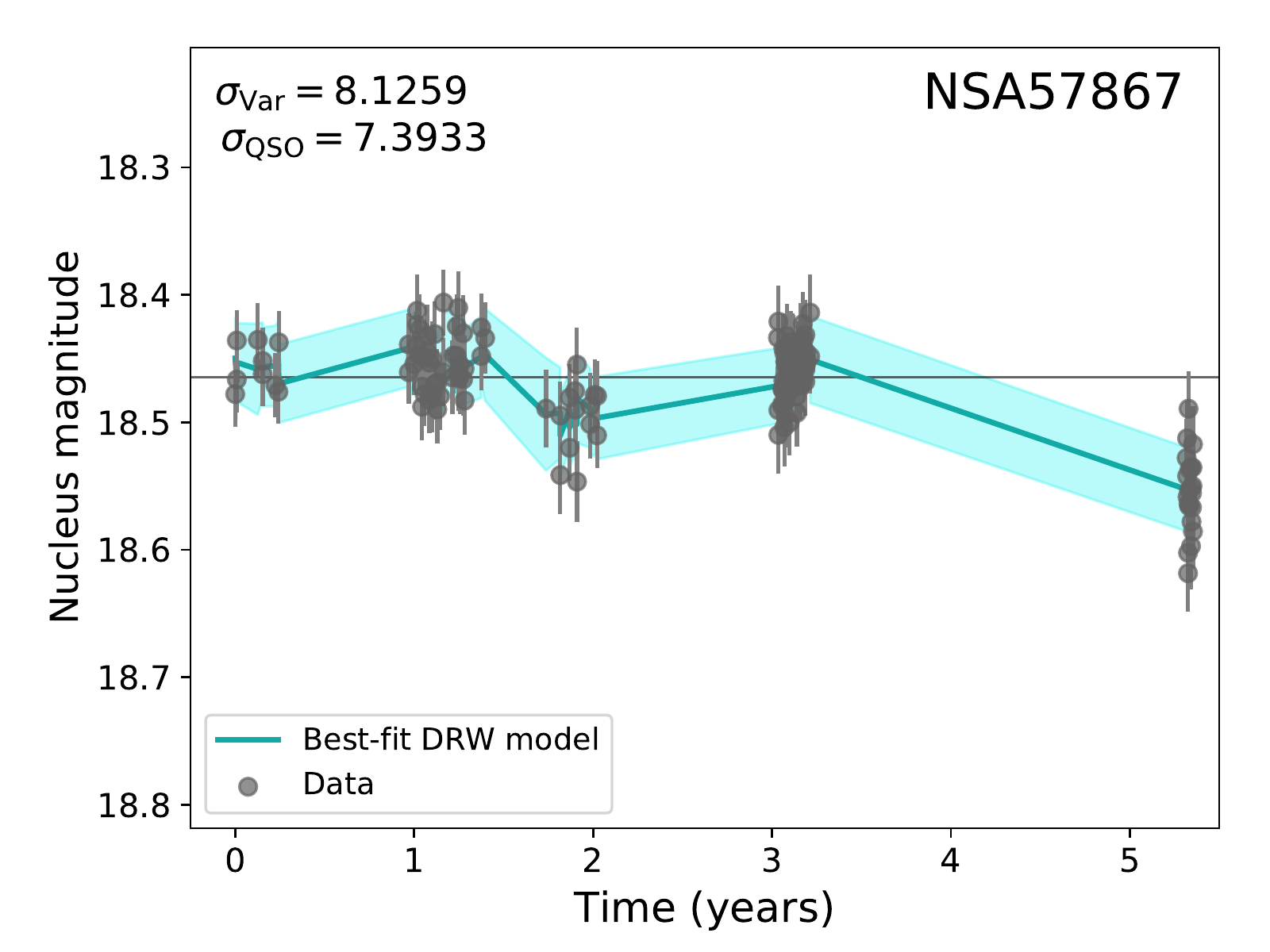}
    \end{minipage}
 
  \begin{minipage}{.26\textwidth}
        \centering
        \includegraphics[width=\textwidth]{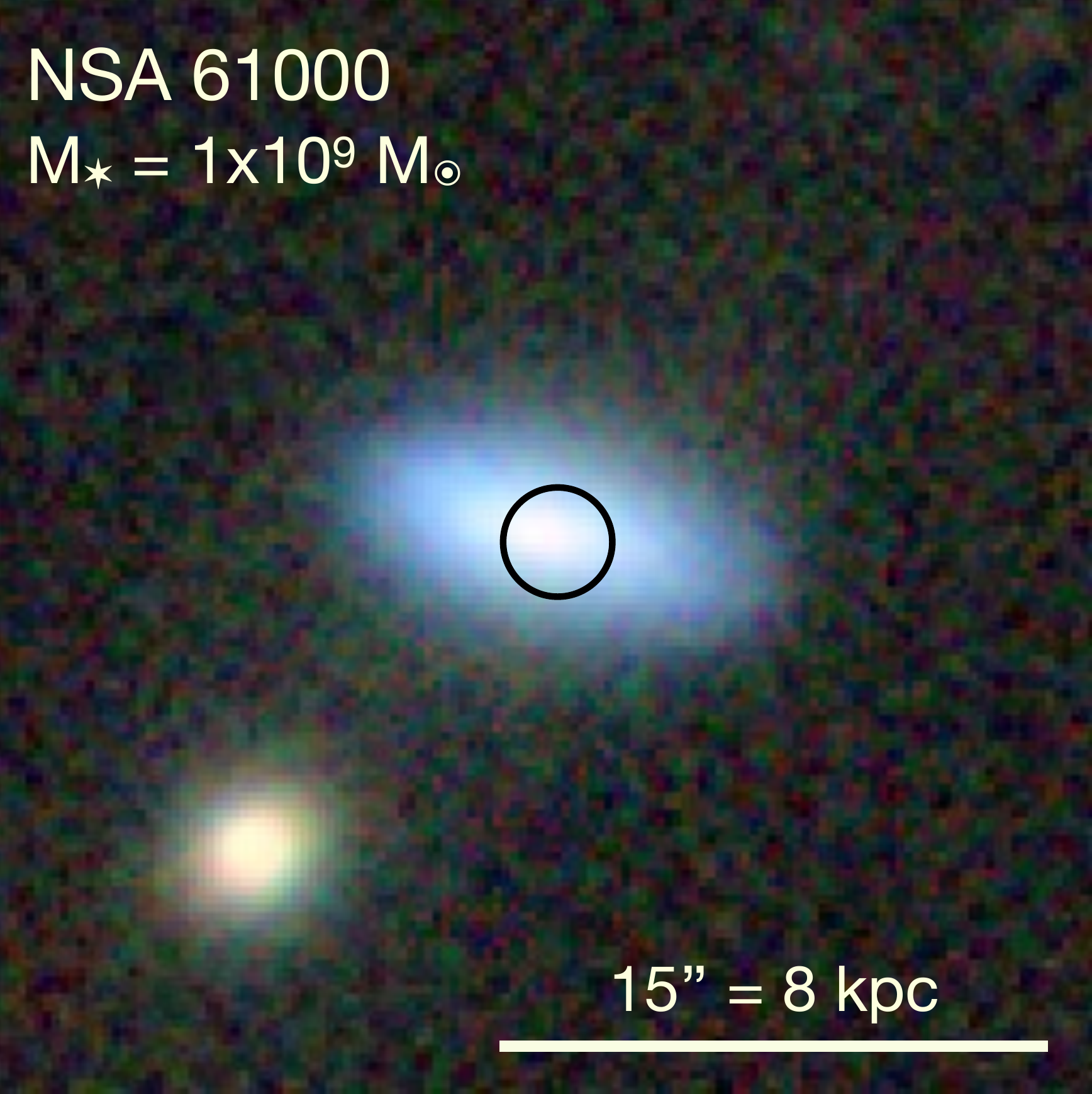}
    \end{minipage}%
    \begin{minipage}{0.55\textwidth}
        \centering
        \includegraphics[width=\textwidth]{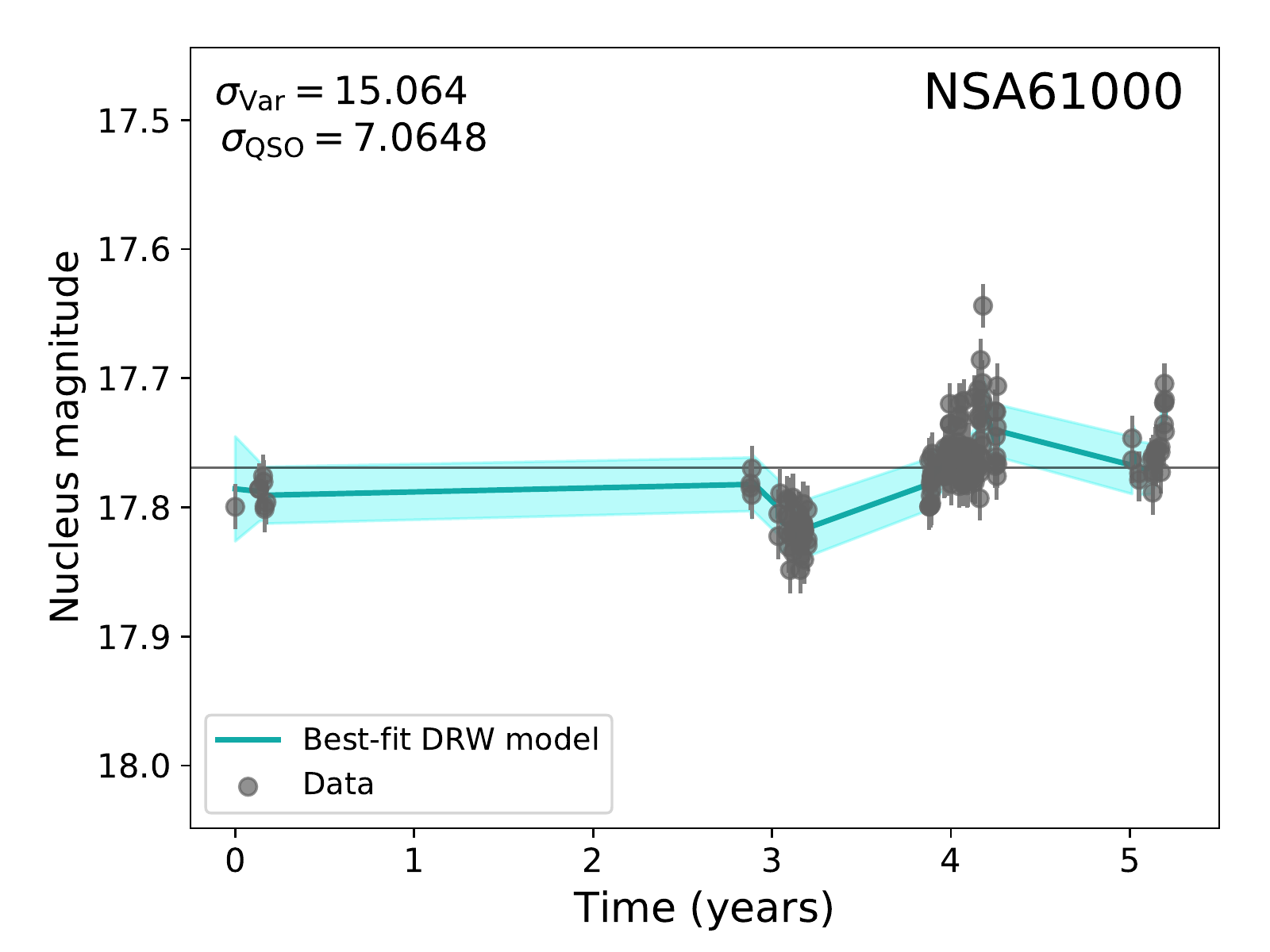}
    \end{minipage}

  \begin{minipage}{.26\textwidth}
        \centering
        \includegraphics[width=\textwidth]{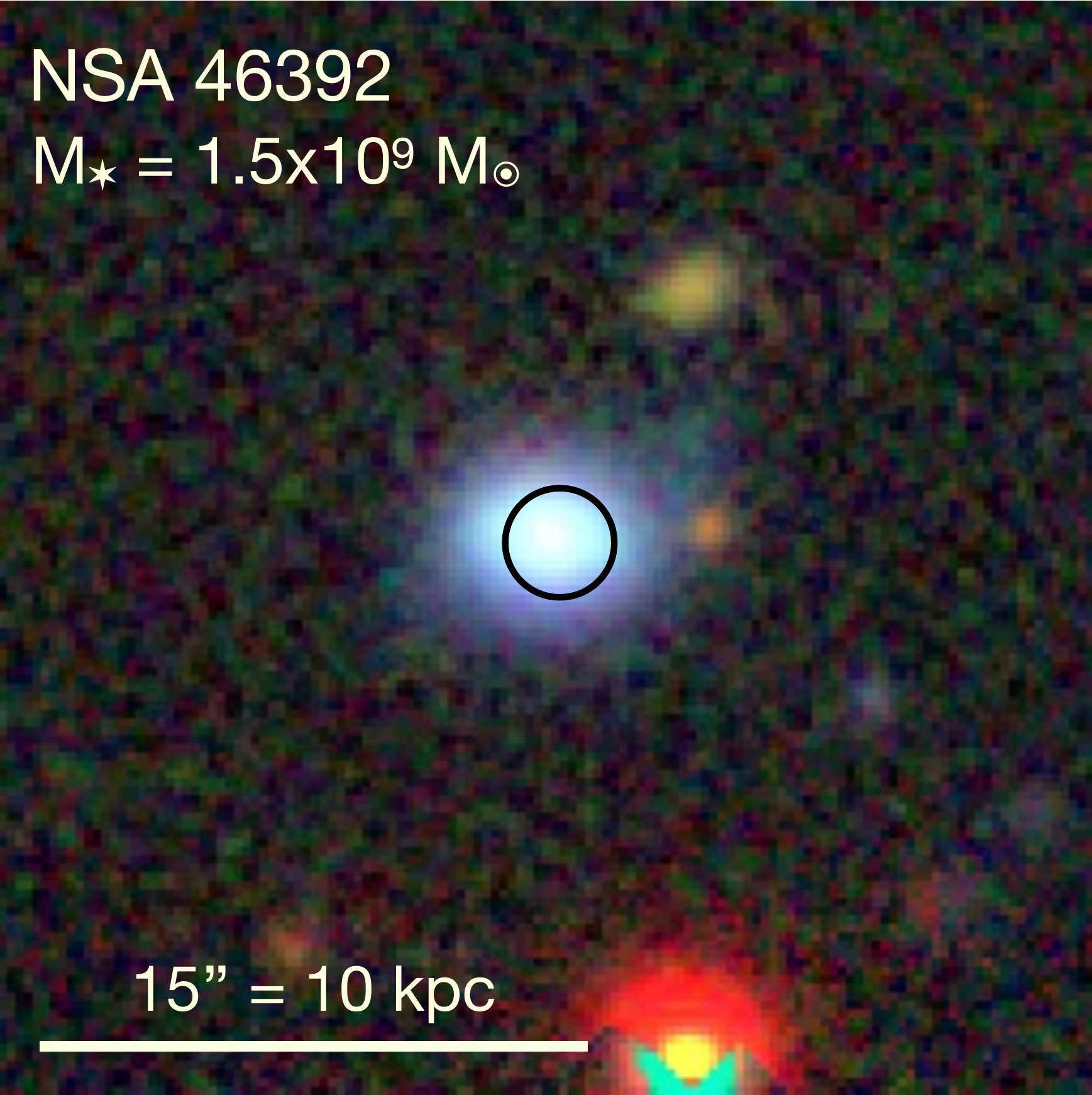}
    \end{minipage}%
    \begin{minipage}{0.55\textwidth}
        \centering
        \includegraphics[width=\textwidth]{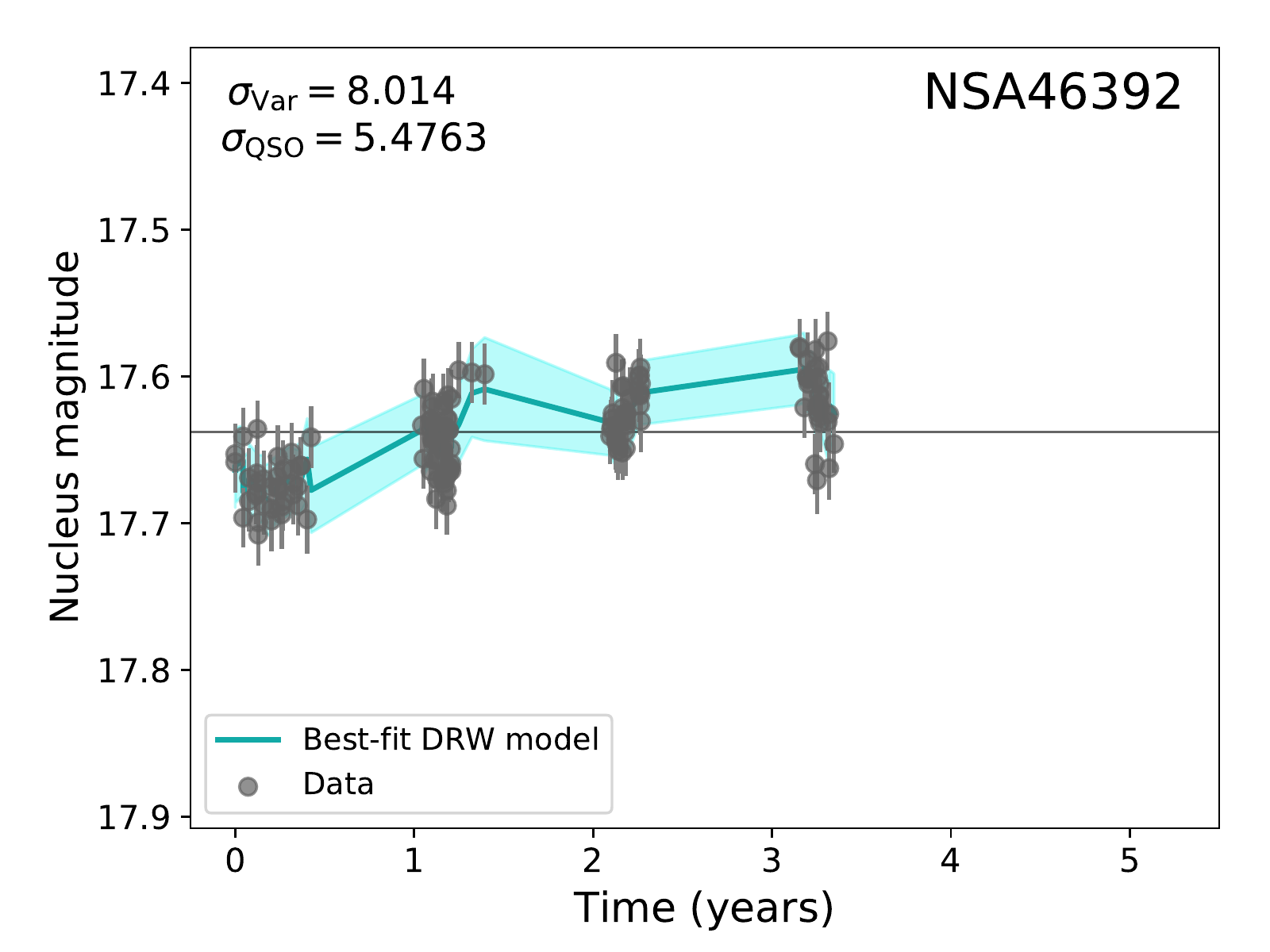}
    \end{minipage}    
    
 \caption{Three low-mass galaxies with star formation-dominated narrow emission lines and AGN-like variability. These galaxies span from $5\times10^{7}-2\times10^{9}~M_{\odot}$ in stellar mass. The images cutouts (left) are from the DECaLS Legacy Survey browser and are 30$''$ on each side. The black circles are the measured PTF R-band magnitude values measured within a 3$''$ nuclear aperture. The light curves (right) give the PTF R-band nucleus magnitudes over 5.5 years. In the light curves, the gray circles are data, the blue line represents the best-fit damped random walk model, and the shaded blue region shows the range of model uncertainties. }
\label{lowmassAGN}
 \end{figure*}   

In this section, we present the main results of our difference imaging and light curve analysis. We compute the fractions of variable AGNs found in the main NSAv0 sample and amongst broad line AGN, and explore how the variable fraction changes as a function of stellar mass and baseline. A summary of our results is given in Table~\ref{tab:fracsummary}. Care should be taken in comparing the fractions from different groups as they may have different magnitude distributions. 

\begin{deluxetable}{c|c|c}   
\tablecaption{Summary of variability percentages \label{tab:fracsummary}}
\tablecolumns{3}
\tablenum{1}
\tablewidth{0pt}
\tablehead{
\colhead{Category} & 
\colhead{${\rm N_{galaxies}}$} & 
\colhead{Variable AGN (\%)} 
}
\startdata
Overall & 47125 & $0.9\pm0.05$ \\
$M_{\ast} >10^{10}~M_{\odot}$ & 12052 &  $1.5\pm0.15$ \\
$M_{\ast} < 10^{10}~M_{\odot}$ & 35073 &  $0.7\pm0.06$ \\
BPT AGN/Comp & 8355 & $1.7\pm0.2$ \\
BPT SF & 28749 &  $0.6\pm0.06$ \\
Broad line AGN & 249 &  $27.7\pm4.0$ \\
Baseline < 2 years & 6854 &  $0.25\pm0.09$ \\
Baseline > 2 years & 40139 &  $1.0\pm0.05$ \\
\enddata
\tablecomments{Variability percentages for different subsamples. These reflect the number of variable AGN in each sub-sample divided by the total number of objects in that sample. The BPT AGN and SF classes are based on the emission line strengths reported in the NSAv0. Uncertainties are binomial limits for a 90\% confidence level \citep{Gehrels:1986kx}. } 
\end{deluxetable}

\subsection{Main NSA sample}
There are \textcolor{black}{78872} galaxies in our main NSAv0 sample with some PTF coverage. Of these, we analyze \textcolor{black}{47125} which have light curves with 20 or more data points. The sample ranges in stellar mass from $\sim10^{7}-10^{12}~M_{\odot}$, though by choice we are only complete for stellar masses below $M_{\ast}=2\times10^{10}~M_{\odot}$. Based on the selection criteria described in Section 3.2, \textcolor{black}{424} out of \textcolor{black}{47125} galaxies have AGN-like variability, for an overall variable AGN fraction of 0.9\%. In Figure~\ref{std_v_mag}, we show the light curve standard deviation versus the PTF R-band magnitude for the full sample. In general, the scatter about the median magnitude for a given light curve increases as the magnitude increases, i.e., there is increased scatter in light curves of fainter objects. The sample of low-mass galaxies from the NSAv0 with AGN-like variability is presented in Table~\ref{tab:PTF_lm_var}. Three examples of low-mass galaxies with AGN-like variability are shown in Figure~\ref{lowmassAGN}.

\begin{figure*}
\centering
\includegraphics[width=0.8\textwidth]{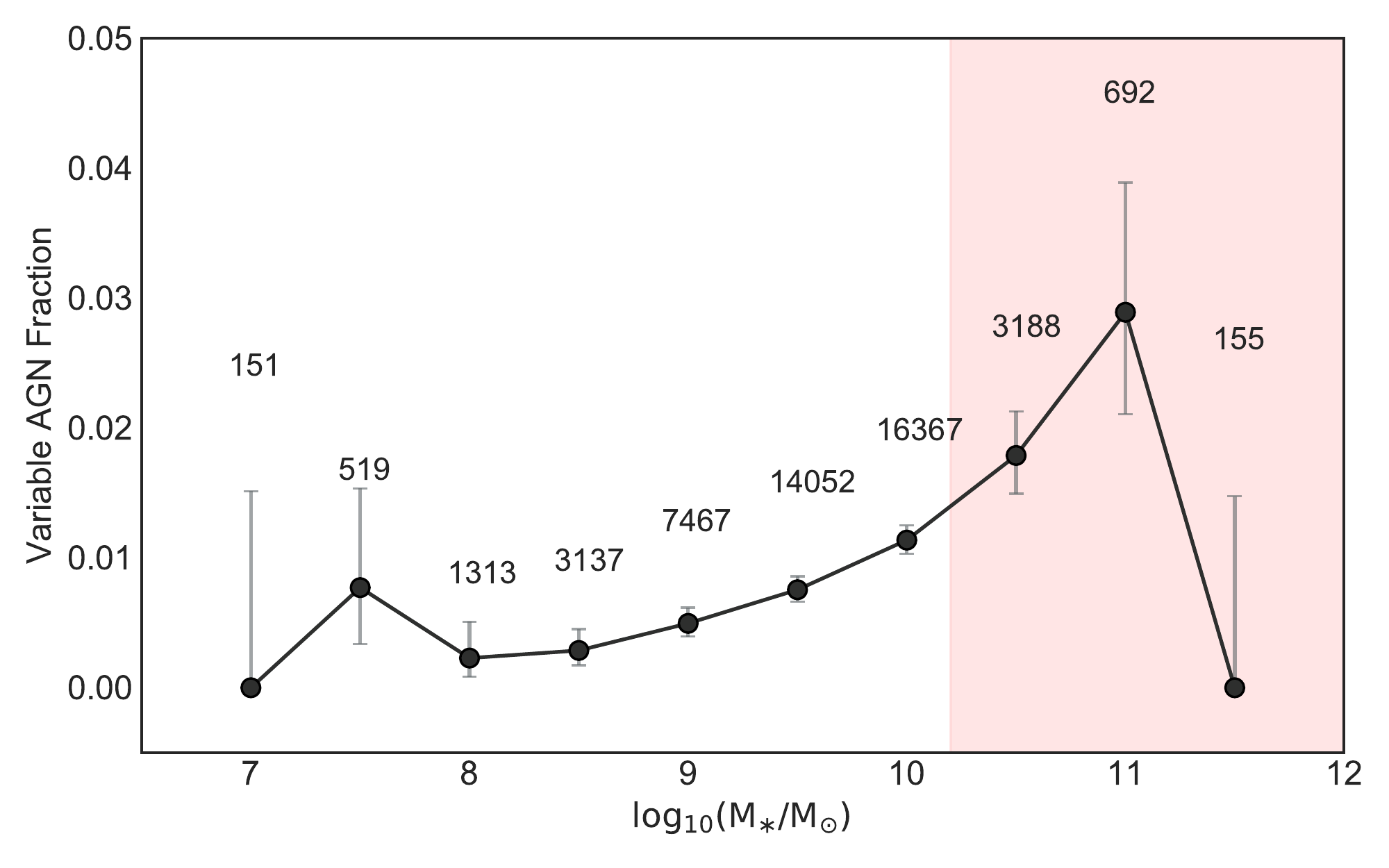}
\caption{Fraction of variability-selected AGN versus galaxy stellar mass. The red shaded region shows the mass range where our sample is incomplete, and the number above each point gives the total number of galaxies in that mass bin. The error bars are the binomial limits for a 90\% confidence level \citep{Gehrels:1986kx}.}
\label{frac_v_mstar}
\end{figure*}

\begin{figure}
\includegraphics[width=0.5\textwidth]{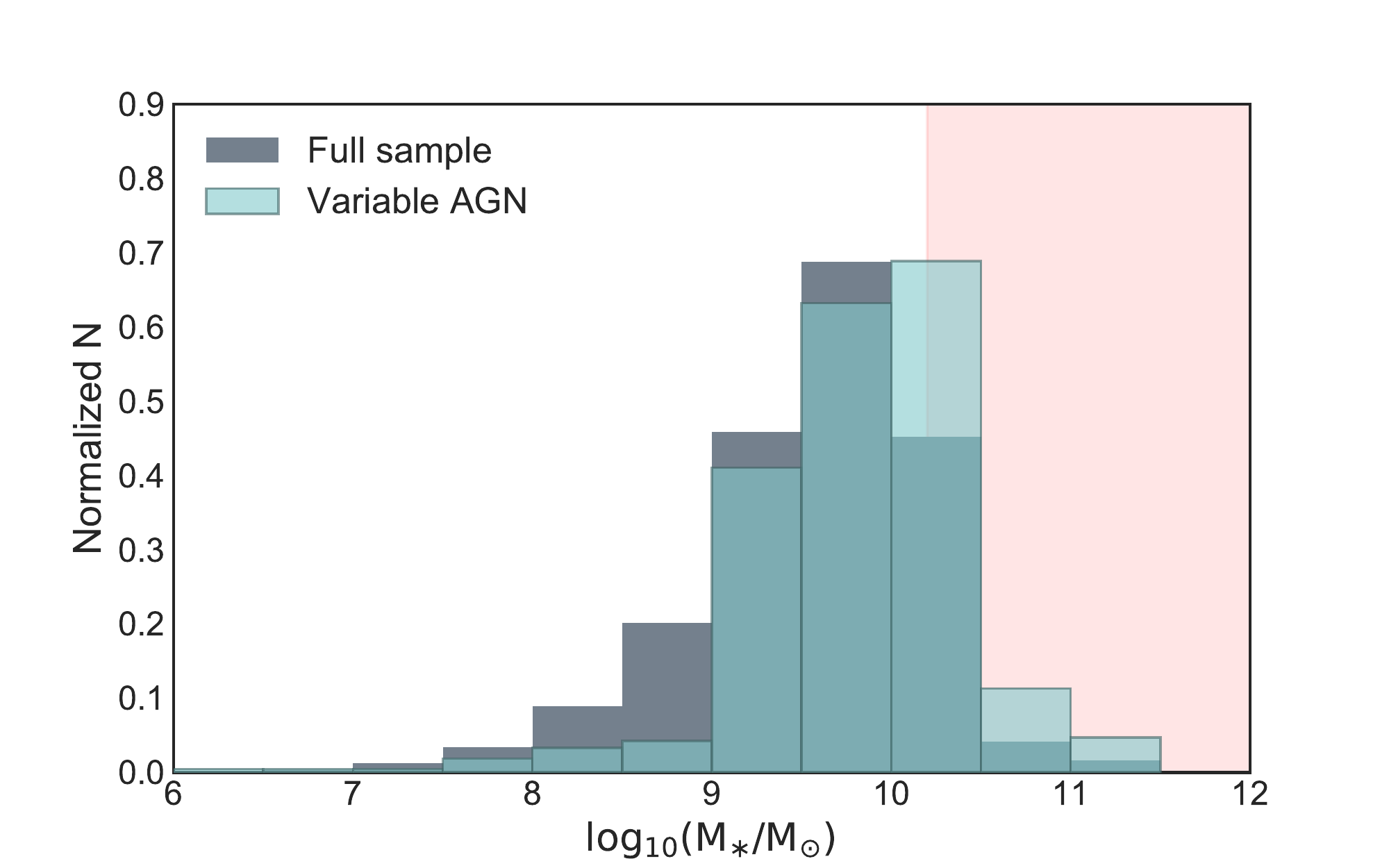}
\caption{Normalized histograms of the stellar mass distribution of the full NSAv0 sample (gray), and galaxies with AGN-like variability (light blue). The red shaded region shows the mass range where our sample is incomplete.  }
\label{masshist}
\end{figure}

We explore the fraction of variable AGN as a function of galaxy and light curve properties. Uncertainties are binomial limits computed following \cite{Gehrels:1986kx}. In Figure~\ref{frac_v_mstar}, we show the fraction of variability selected AGN as a function of host galaxy stellar mass; we find that the AGN fraction increases towards higher stellar masses. In Figure~\ref{masshist}, we show histograms of stellar mass for the full sample and for variable AGN. The sample of variable AGN skews towards higher stellar masses; the median stellar mass of the overall sample is $4.5\times10^{9}~M_{\odot}$, while the median stellar mass of the variability-selected AGN is $8.3\times10^{9}~M_{\odot}$. The pink shaded region in Figures~\ref{frac_v_mstar} and~\ref{masshist} denote the region $M_{\ast}>2\times10^{10}~M_{\odot}$ where our sample is incomplete in stellar mass.

\begin{figure}
\includegraphics[width=0.5\textwidth]{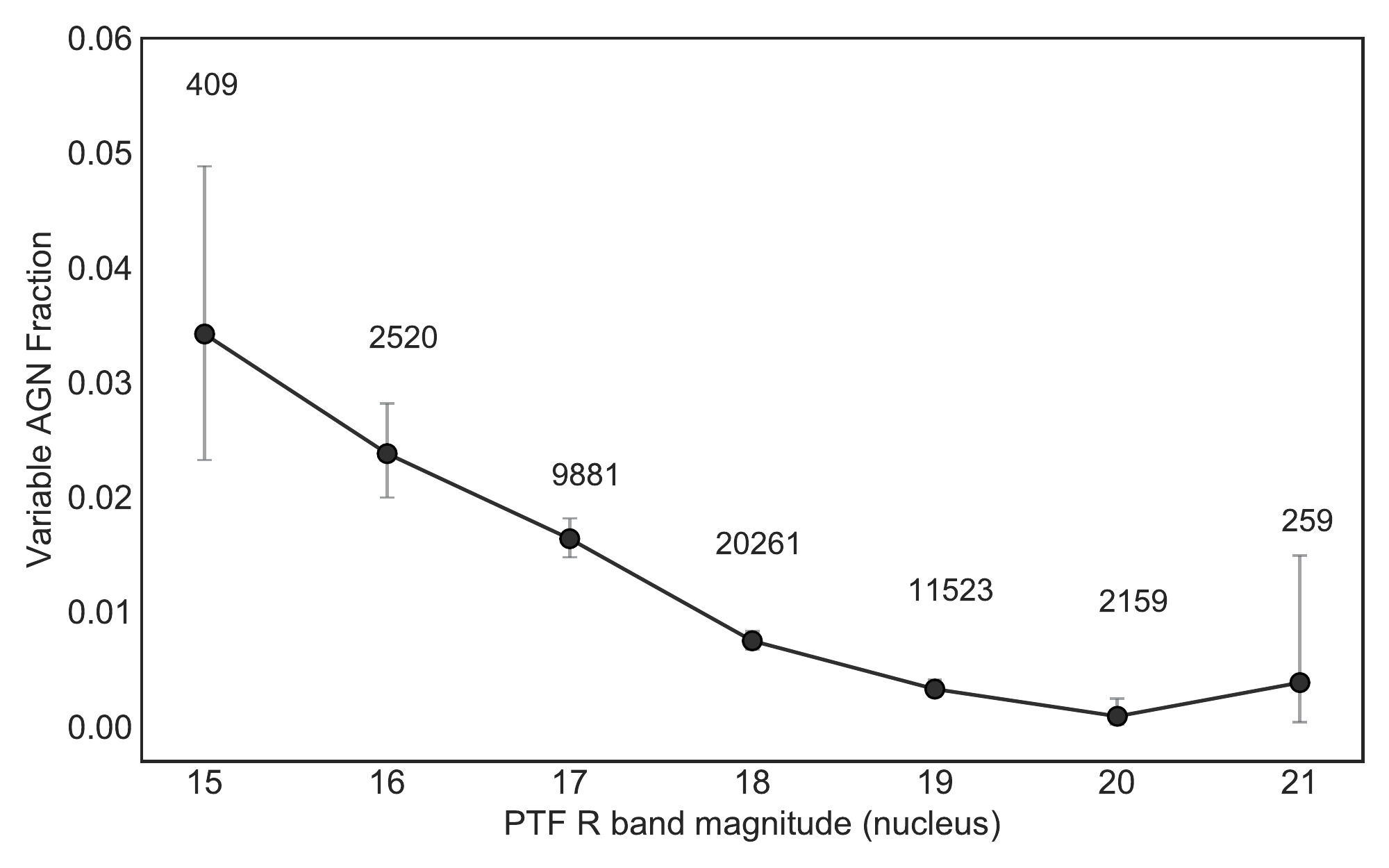}
\includegraphics[width=0.5\textwidth]{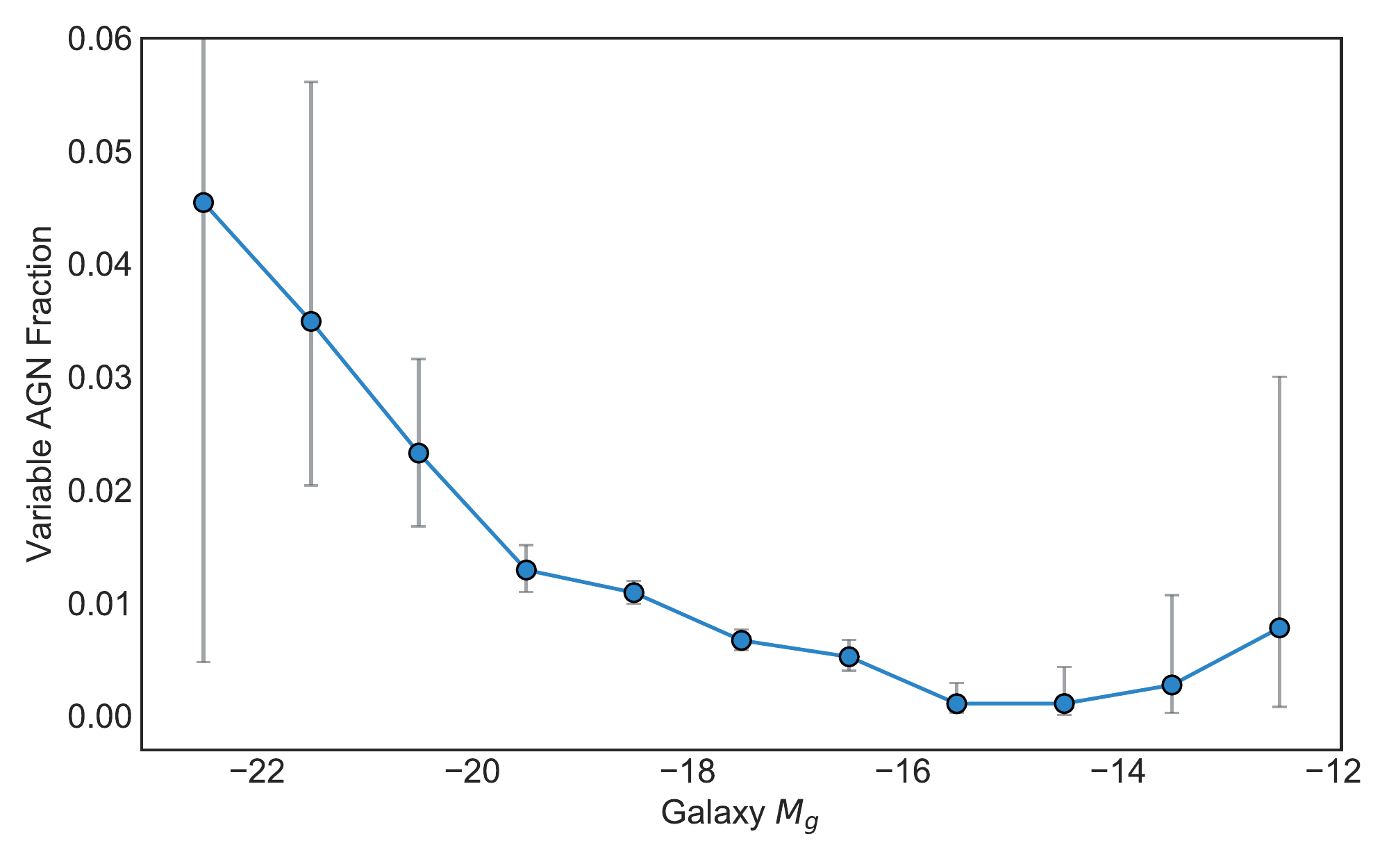}
\caption{Top: the fraction of variable AGN versus the nucleus PTF R-band magnitude. Bottom: Fraction of variable AGN versus galaxy absolute g-band  magnitude from the NSAv0.}
\label{AGNfracs_mag}
\end{figure}

The decline in variability fraction towards low stellar masses cannot be taken at face value. As we show in the top panel of Figure~\ref{AGNfracs_mag}, the AGN fraction is also a strong function of the nucleus apparent magnitude. Additionally, Figure~\ref{std_v_mag} showed that the scatter in a given light curve increases as the nucleus magnitude gets fainter, so the lower variability fraction for lower mass galaxies could be due to the fact that the lower mass galaxies are fainter. We address this further in Section 6.

\begin{figure*}
\centering
\includegraphics[width=0.7\textwidth]{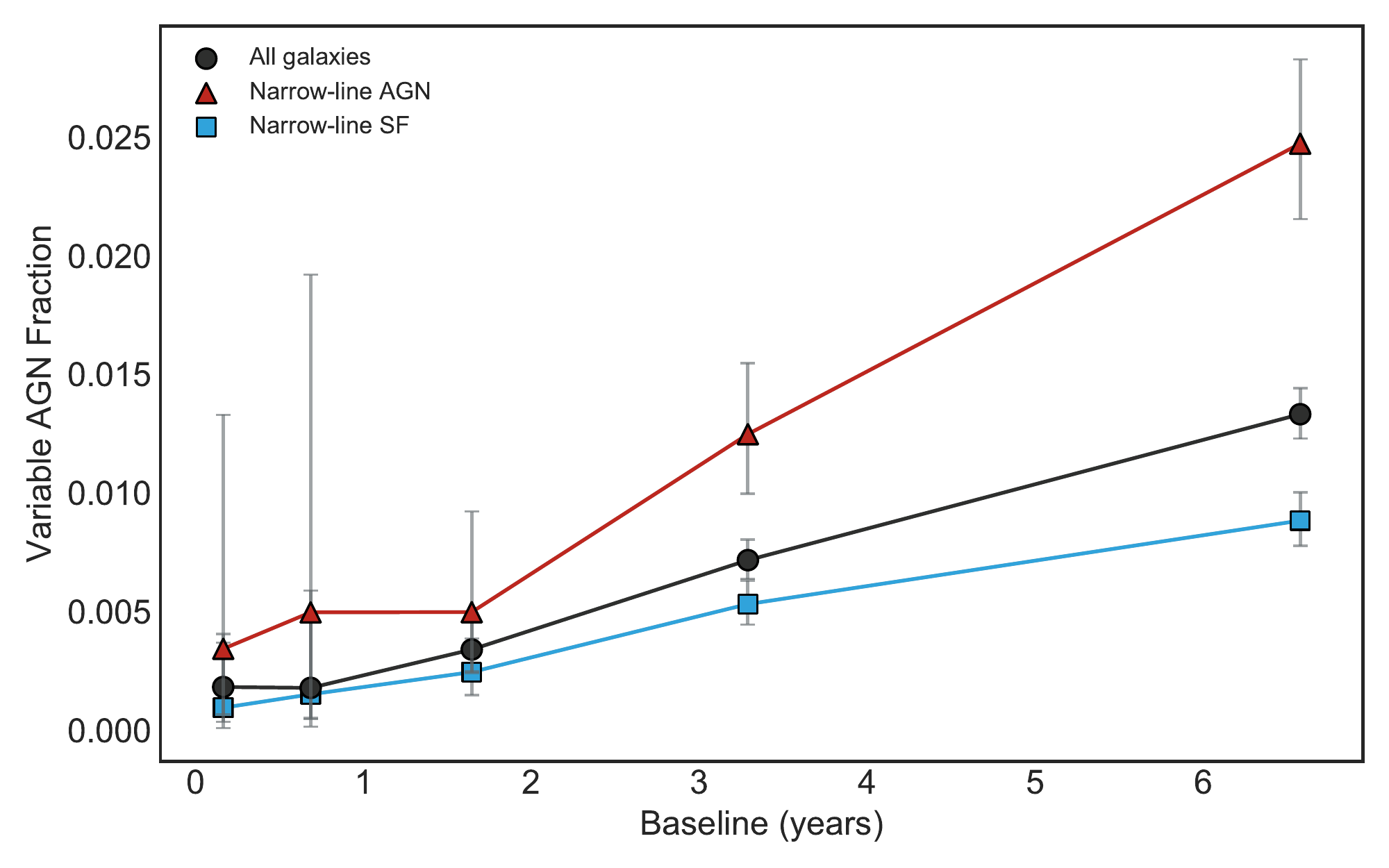}
\caption{Variable AGN fraction versus baseline for the NSAv0 sample. The black circles show the variable AGN fraction for the entire sample. The red triangles and blue squares show the fraction for narrow-line AGN and star forming galaxies, respectively.  }
\label{AGNfrac_v_baseline}
\end{figure*}

In Figure~\ref{AGNfrac_v_baseline}, we show the fraction of variable AGN as a function of the light curve baseline, for the full sample, as well as for narrow-line AGN and narrow-line star forming objects. For the overall sample, the fraction of variable AGN increases from $0.25\%\pm0.09\%$ for light curves shorter than 2 years to $1\%\pm0.05\%$ for light curves longer than 2 years. For narrow line AGN, the variable fraction reaches 2.5\% for baselines longer than 6 years. Especially as data from LSST begins to come in, any estimate of the variable AGN fraction for a population should take into account the baseline over which it is being measured.

\begin{deluxetable*}{cccccccccccc} 
\tablecaption{ 244 Low-mass galaxies with AGN-like variability\label{tab:PTF_lm_var}}
\tablecolumns{12}
\tablenum{2}
\tablewidth{0pt}
\tabletypesize{\footnotesize}
\tablehead{
\colhead{NSA ID } &
\colhead{RA} &
\colhead{Dec } &
\colhead{Redshift} &
\colhead{Stellar mass} &
 \colhead{N$_{\rm points}$ } & 
 \colhead{Baseline} &  
\colhead{Median $m_{R}$}  &
\colhead{Std. dev.} & 
\colhead{$\sigma_{\rm var}$ } &
\colhead{$\sigma_{\rm QSO}$ } &
\colhead{$\sigma_{\rm notQSO}$ } \\
\colhead{ } &
\colhead{} &
\colhead{ } &
\colhead{} &
\colhead{ [$\log_{10}(M_{\odot})$] } &
 \colhead{ } & 
 \colhead{[days]} &  
\colhead{[mag]}  & 
\colhead{[mag]} & 
\colhead{ } &
\colhead{ } &
\colhead{ }
}
\startdata
NSA126546 & 4.29825 & 29.4809 & 0.0235 & 9.99 & 209 & 1546.75 & 17.16 & 0.014 & 2.65 & 2.47 & 0.04\\
NSA126631 & 4.65217 & 29.93285 & 0.0215 & 9.84 & 154 & 1546.75 & 17.68 & 0.024 & 6.52 & 4.09 & 0.18\\
NSA23643 & 10.19812 & 0.62188 & 0.0414 & 9.84 & 153 & 1863.94 & 17.58 & 0.019 & 2.8 & 2.34 & 0.11\\
NSA5956 & 13.79728 & -0.77002 & 0.0425 & 9.98 & 107 & 1551.92 & 17.45 & 0.026 & 7.96 & 3.31 & 1.19 \\
NSA128247 & 14.97201 & 31.82702 & 0.0149 & 9.9 & 59 & 1540.92 & 15.3 & 0.125 & 166.16 & 10.86 & 8.73\\
NSA44590 & 16.44724 & 0.50798 & 0.05 & 9.33 & 131 & 1551.88 & 18.83 & 0.041 & 4.08 & 2.08 & 0.85\\
NSA62320 & 20.6387 & 0.13574 & 0.0441 & 9.15 & 282 & 1506.02 & 19.46 & 0.059 & 8.21 & 3.62 & 0.9\\
NSA129568 & 21.40065 & 34.73646 & 0.0162 & 9.83 & 264 & 1458.15 & 16.2 & 0.012 & 7.03 & 6.45 & 0.0\\
NSA8747 & 22.22581 & 14.9397 & 0.0332 & 9.34 & 23 & 1180.06 & 18.19 & 0.047 & 6.46 & 2.67 & 0.96\\
NSA9053 & 27.32894 & 13.04779 & 0.0178 & 9.72 & 238 & 1867.07 & 16.93 & 0.016 & 8.74 & 8.04 & 0.0\\
NSA44431 & 27.4215 & 0.89644 & 0.0284 & 8.84 & 49 & 1533.73 & 19.48 & 0.058 & 2.55 & 2.63 & 0.11\\
NSA9177 & 28.23465 & 14.49017 & 0.0438 & 9.73 & 233 & 1883.93 & 17.73 & 0.022 & 2.75 & 3.33 & 0.02\\
NSA7119 & 34.89554 & -0.40896 & 0.0259 & 8.67 & 79 & 1551.91 & 18.17 & 0.029 & 2.91 & 3.74 & 0.01 \\
NSA7537 & 41.06829 & 0.3416 & 0.0221 & 9.74 & 293 & 1540.88 & 17.67 & 0.022 & 9.51 & 2.49 & 2.29 \\
NSA7573 & 43.12908 & 0.20657 & 0.0522 & 9.74 & 286 & 1655.68 & 19.95 & 0.123 & 9.71 & 6.49 & 1.41 \\
NSA62560 & 44.28604 & -0.47988 & 0.0379 & 9.38 & 101 & 1551.89 & 19.02 & 0.073 & 16.68 & 4.64 & 2.62\\
NSA11588 & 55.87608 & -7.58542 & 0.0357 & 9.84 & 21 & 1027.17 & 17.79 & 0.036 & 4.87 & 3.88 & 0.1 \\
NSA171680 & 57.34663 & -11.99095 & 0.0321 & 9.36 & 25 & 711.1 & 17.91 & 0.053 & 8.66 & 3.61 & 0.65\\
NSA171487 & 68.17618 & -4.38251 & 0.0149 & 9.43 & 152 & 1180.03 & 17.09 & 0.026 & 18.83 & 5.16 & 3.12\\
NSA109532 & 119.84415 & 9.61993 & 0.0086 & 9.75 & 89 & 1475.96 & 16.49 & 0.011 & 5.01 & 3.02 & 0.25\\
... & ... & ... & ... & ... & ... & ... & ... & ... & ... & ...& ... \\
\enddata
\tablecomments{Variability-selected AGN with $M_{\ast}<10^{10}~M_{\odot}$, in order of Right Ascension. NSA ID refers to the ID given in the NSAv0. $\sigma_{\rm var}$, $\sigma_{\rm QSO}$, and $\sigma_{\rm notQSO}$ refer to the significance that the given light curve is variable, an AGN, or a false alarm, respectively. Details on these parameters can be found in Section 3.2. A full version of this table is available in the online version.  }
\end{deluxetable*}

\subsubsection{Spectroscopic properties}

\begin{figure*}
\centering
\includegraphics[width=0.9\textwidth]{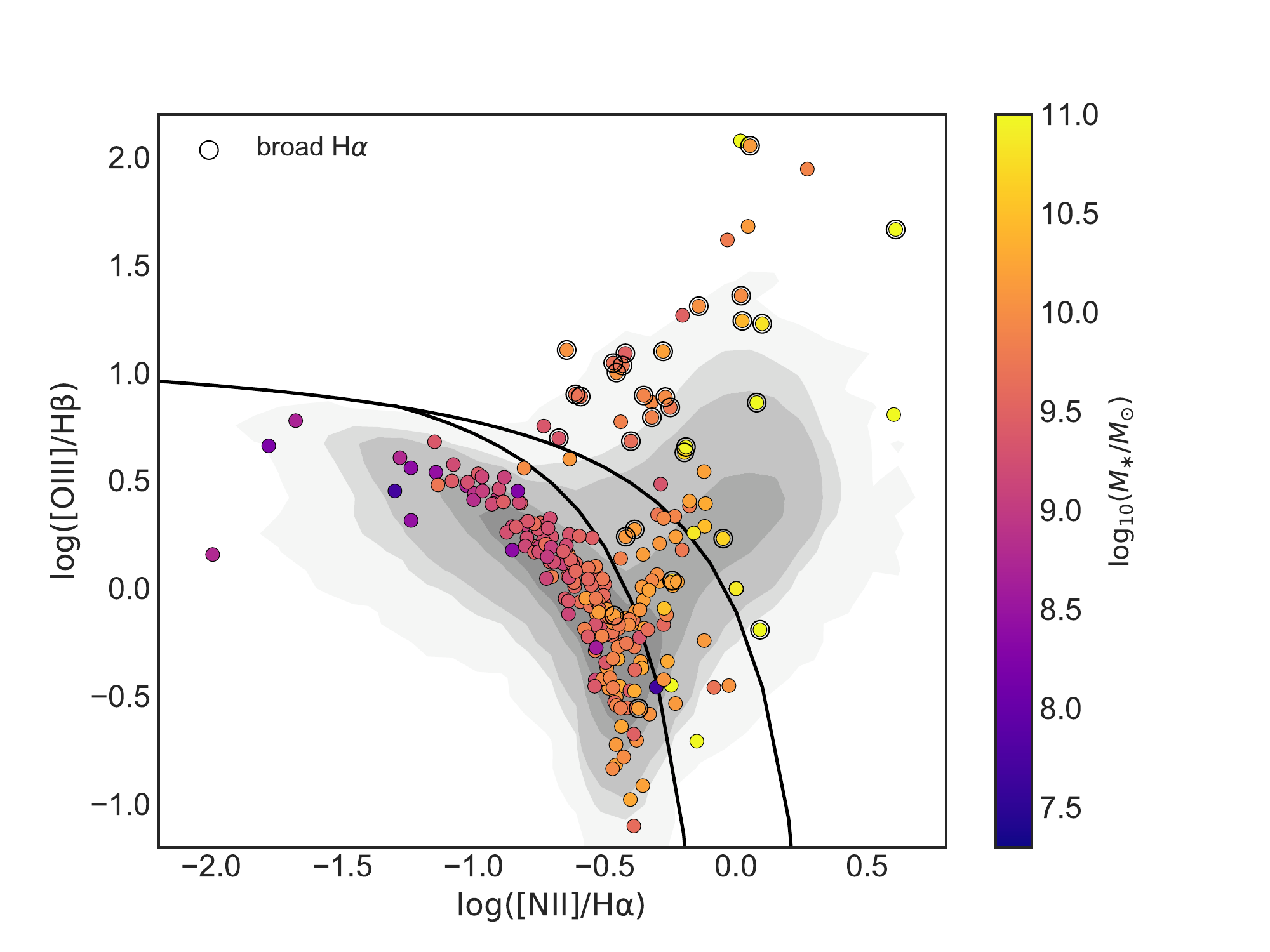}
\caption{BPT diagram of galaxies with AGN-like variability. Each filled circle represents a galaxy, and the points are color coded by galaxy total stellar mass. Points with a circle around them indicate the presence of broad H$\alpha$ emission. Most low-mass galaxies reside in the star-forming (lower left) region of the BPT diagram.} 
\label{BPT}
\end{figure*}

SDSS spectra are available for \textcolor{black}{357} of the \textcolor{black}{424} galaxies with AGN-like variability. The galaxies without SDSS spectra have spectroscopic redshifts available from other databases, and we do not analyze their spectra here. In Figure \ref{BPT}, we show where the variability selected AGN from the NSAv0 fall on the BPT diagram. 84 galaxies could not be placed on the BPT diagram as one or more of the relevant lines were absorption dominated. Of the 273 galaxies on the BPT diagram, $40\%$ are in the AGN/composite regions, and $60\%$ are in the star forming region. 

The high fraction of variable AGN in the star forming region is driven by the fact that our sample is dominated by low-mass galaxies. Among galaxies with $M_{\ast}>10^{10}~M_{\odot}$, $70\%$ are in the AGN/composite regions; for galaxies with $M_{\ast}<10^{10}~M_{\odot}$, $25\%$ are in the AGN/composite region. There are also 35 galaxies with broad H$\alpha$ emission, roughly two-thirds of which have $M_{\ast}>10^{10}~M_{\odot}$. Two of the 35 broad H$\alpha$ galaxies fall in the star forming region of the BPT diagram; the remainder are in the composite or AGN regions. 

Galaxies with AGN could fall in the star forming region of the BPT diagram due to star formation dilution of the AGN signal within the SDSS spectroscopic fiber or due to low-metallicity. As shown in \cite{2006MNRAS.371.1559G}, galaxies with sub-solar metallicity would lie in the upper left region of the diagram ($\log_{10}\rm{([NII]/H{\alpha})} < -1.0$ and $\log_{10}\rm{([OIII]/H{\beta})}$ ranging from $\sim0.25-1.0$). We find a handful of galaxies in this regime; these tend to be some of the lowest mass galaxies in our variable sample. The fraction of galaxies with AGN-like variability with SDSS optical spectroscopy dominated by star formation demonstrates that particularly at low galaxy stellar masses, AGN selection via photometric variability can find AGN that would be missed by standard optical spectroscopic selection techniques. 

\subsection{Additional samples}
Here we present variability results for our ancillary samples: broad-line AGN from \cite{2007ApJ...670...92G}, \cite{2015ApJ...813...82R}, variability-selected AGN from \cite{2018ApJ..868..152}. 

\subsubsection{Broad-line AGN} 

In order to explore trends between variability properties and black hole mass, we also construct light curves for broad line AGN from \cite{2007ApJ...670...92G} and \cite{2015ApJ...813...82R}. The objects from \cite{2007ApJ...670...92G} are generally at higher redshift than the NSAv0 and so are not in our main sample. While \cite{2015ApJ...813...82R} also uses the NSAv0 as their parent sample, not all of their high-mass galaxies end up in our main sample.

\cite{2007ApJ...670...92G} searched the SDSS DR7 for galaxies with broad H$\alpha$ emission indicative of BH with $M_{\rm BH}\lesssim10^{6}~M_{\odot}$ and identified 229 galaxies meeting this criterion. \cite{2015ApJ...813...82R} searched for broad H$\alpha$ emission in 66,945 galaxies from the NSAv0. Their final sample was composed of 244 broad-line AGN with BH masses ranging from $M_{\rm BH}\approx10^{5}-10^{8.5}~M_{\odot}$. Of these 473 broad-line AGN, we can construct light curves with more than 20 data points for 192. There are an additional 57 broad-line AGN within z<0.055 from the \cite{2019ApJS..243...21L} catalog of broad-line AGN in the SDSS DR7 that have light curves in our PTF sample. This gives a final count of 249 broad-line AGN for which we analyze PTF light curves. Overall, we find that 69/249 objects show AGN-like variability ($28\%$). 

\begin{figure*}
\centering
\includegraphics[width=\textwidth]{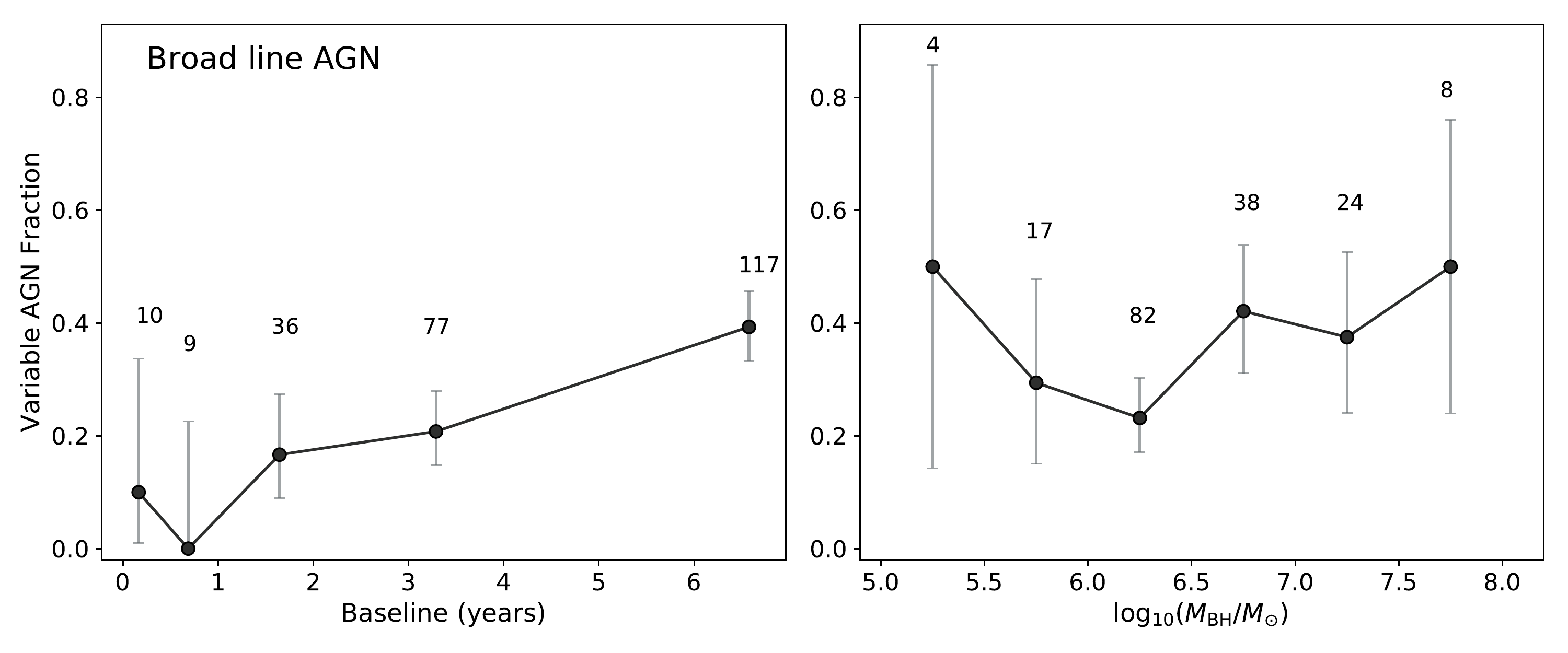}
\caption{Variable fraction versus baseline (left) and BH mass (right) for broad-line AGN from \cite{2007ApJ...670...92G}, \cite{2015ApJ...813...82R}, and \cite{2019ApJS..243...21L}. The variability fraction for broad line AGN increases to 40\% for baselines of 6 years. We find no change in the variability fraction as a function of BH mass.   }
\label{Varfrac_GHRV}
\end{figure*}

In Figure~\ref{Varfrac_GHRV}, we show the variable fraction as a function of baseline. For baselines of 2500 days, almost 40\% of broad-line AGN are found to be variable, compared to $\sim20\%$ for light curve baselines of $\sim500$ days. We also plot the variability fraction as a function of the BH mass, and find that the fraction remains roughly constant with BH mass. This suggests that, at least down to BH masses of $\sim10^{5}~M_{\odot}$, the presence of AGN variability is not dependent on the mass of the central BH.

\subsubsection{Comparison to Stripe 82 variability-selected AGN}

In \cite{2018ApJ..868..152}, we constructed light curves of $\sim28,000$ galaxy nuclei using SDSS Stripe 82 data. The full sample ranged in stellar mass from $M_{\ast}\approx 10^{7}-10^{12}~M_{\odot}$. Of the 28,062 galaxies, we found 135 with AGN-like variability, 35 of which had $M_{\ast}<10^{10}~M_{\odot}$. We constructed PTF light curves with $\ge20$ data points for 70/135. The baselines for the PTF light curves range from 65 days to 1903 days. Of the 70 \cite{2018ApJ..868..152} galaxies with PTF light curves, 18 show AGN-like variability in PTF (25\%). 

\begin{figure*}
\centering
\includegraphics[width=0.9\textwidth]{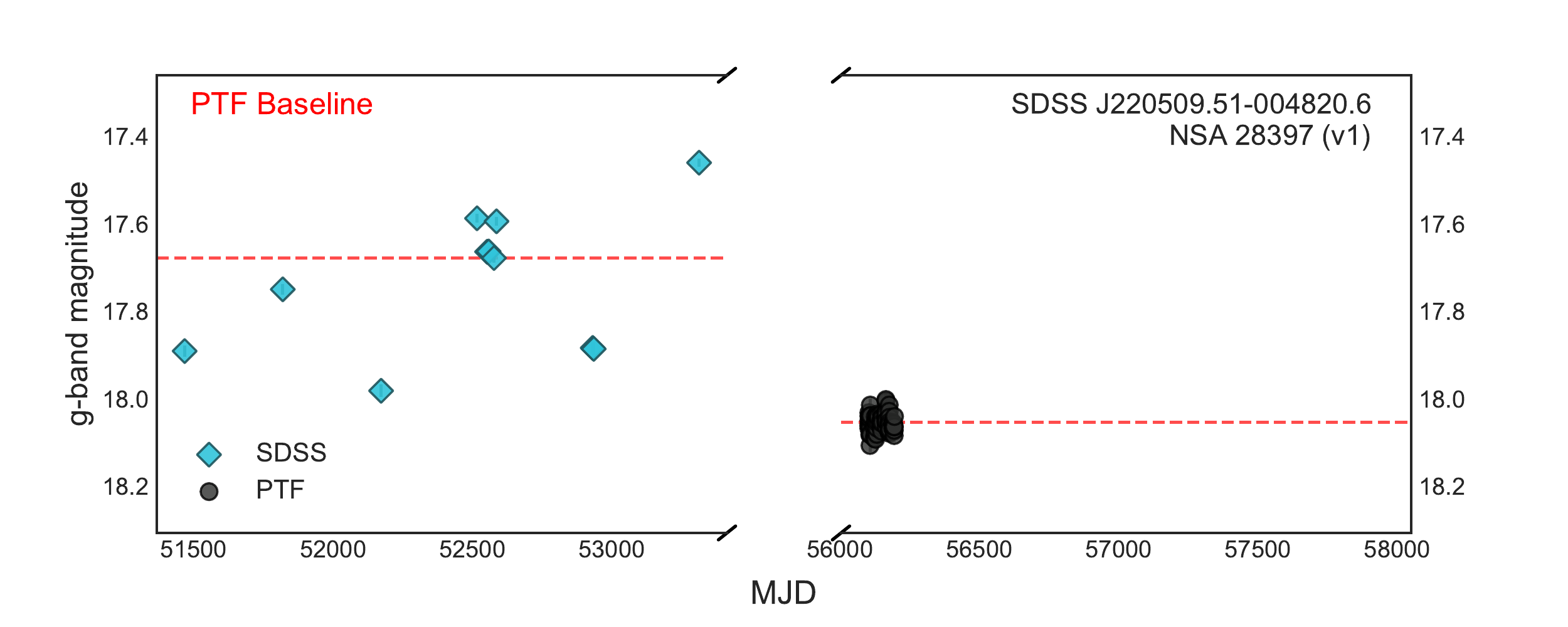}\\
\includegraphics[width=0.9\textwidth]{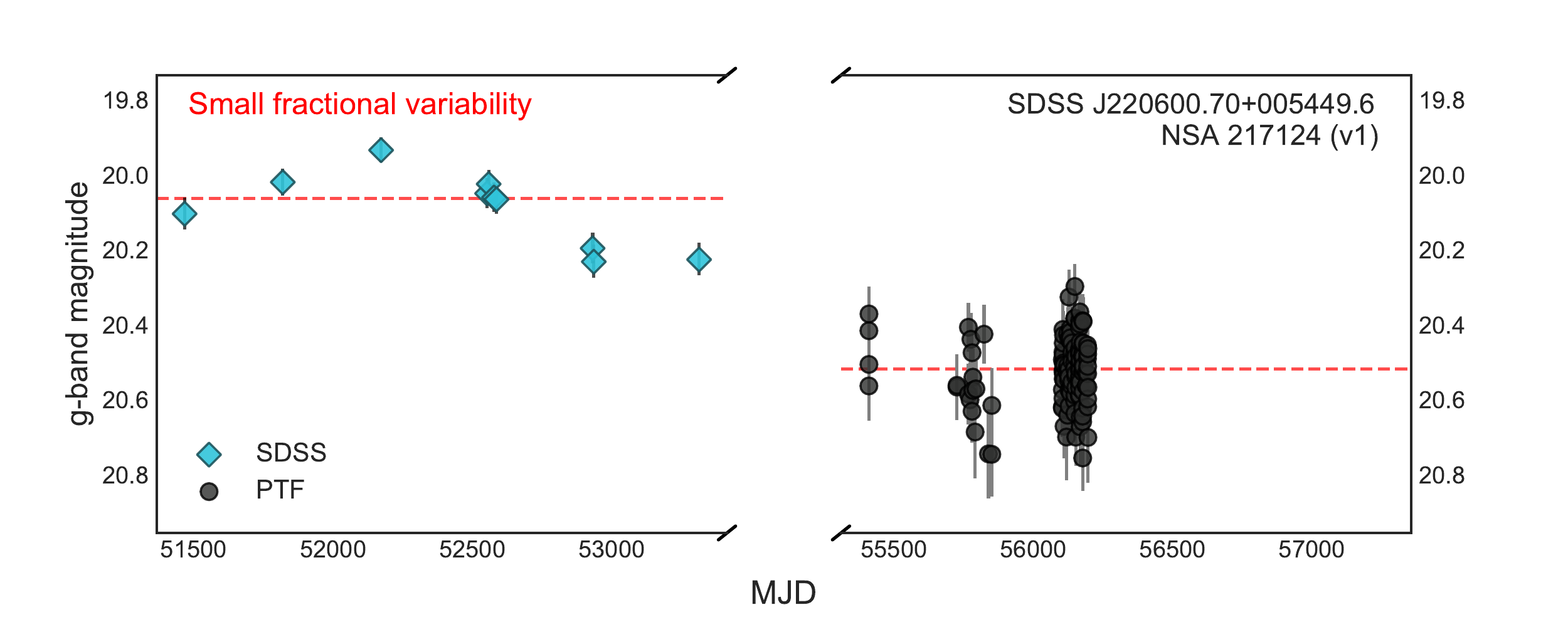}\\
\caption{Light curves for two objects found to be variable AGN in \cite{2018ApJ..868..152} based on Stripe 82 data but not classified as variable AGN based on their PTF light curves. For each object, the SDSS g-band light curve is shown on the left, and the PTF light curve (transformed into g-band magnitudes) is shown on the right. NSA 28397 (top) is not classified as variable in PTF, but has a PTF light curve spanning just 91 days. NSA 217214 (bottom) has a fractional variability in the SDSS light curve which is smaller than the fractional uncertainty on the PTF data points. Note that the photometric transformation from PTF R-band to SDSS g-band (\citealt{2012PASP..124...62O}), is intended for objects with stellar-like spectra.  }
\label{s82_ptf_samp}
\end{figure*}

We identify several reasons for galaxies being identified as variable in SDSS and not in PTF. These results are summarized in Table~\ref{tab:PTFstripecomp}. The first is sampling of the PTF light curve, i.e., the total baseline is too short, or points are clumped at either end of a long baseline with nothing in between. The median Stripe 82 baseline for this sample is 2239 days, while the median PTF baseline for this sample is 1214 days. 16/52 galaxies fall into this category. There are also galaxies where the fractional variations in the SDSS light curve are smaller than the error bars on the corresponding PTF light curve. This is true for 14/52 galaxies. Figure~\ref{s82_ptf_samp} gives examples of these. 

\begin{deluxetable}{c|cccc}   
\tablecaption{Variable Stripe 82 systems in PTF \label{tab:PTFstripecomp}}
\tablecolumns{5}
\tablenum{3}
\tablewidth{0pt}
\tablehead{
\colhead{Variable} & 
\multicolumn4c{Not Variable} 
\\
\colhead{} &
\colhead{Sampling} &
\colhead{Noise} &
\colhead{Changing look } & 
\colhead{Uncertain} 
}
\startdata
18  & 16 & 14 & 8 & 14 \\
\enddata
\tablecomments{Summary of the PTF variability properties of galaxies identified as variable in Stripe 82 \citep{2018ApJ..868..152}. There were 70 galaxies for which we constructed light curves in both Stripe 82 and PTF.} 
\end{deluxetable}

\begin{figure*}
\centering
\includegraphics[width=0.9\textwidth]{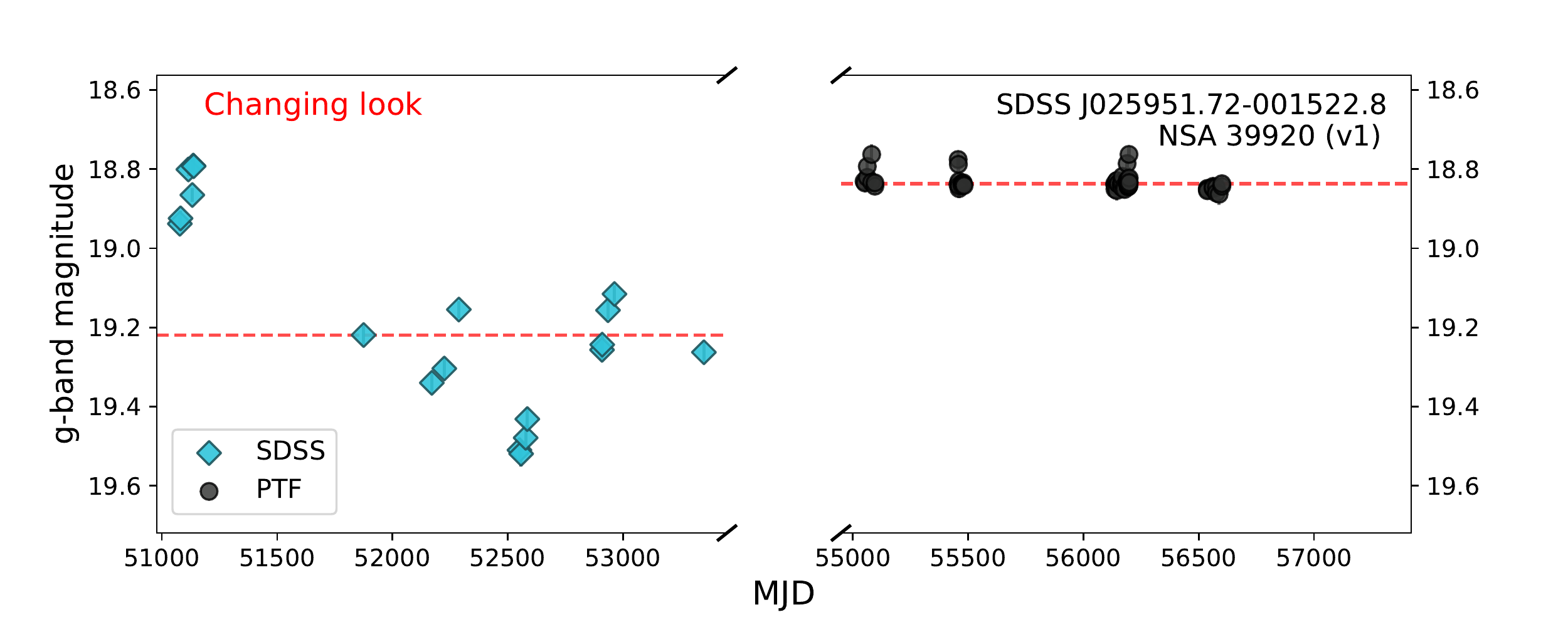}\\
\caption{Light curve for NSA 39920, a broad-like AGN found to be variable in \cite{2018ApJ..868..152} but not in PTF. The SDSS g-band light curve is shown on the left, and the PTF light curve (transformed into g-band magnitudes) is shown on the right. The PTF light curve is well-sampled; if there were variability on the level of the SDSS light curve, it would be observable with PTF.  Note that while it appears that the galaxy is brighter in the PTF light curve, the photometric transformation from PTF R-band to SDSS g-band (given in \citealt{2012PASP..124...62O}), is intended for objects with stellar-like spectra and thus may not provide an accurate transformation here.  }
\label{s82_ptf_cl}
\end{figure*}

We also identify a sample of galaxies which may be ``changing look AGN". Changing look AGN are characterized by the sudden and drastic disappearance or appearance of broad optical emission lines on timescales of years \citep{2014ApJ...788...48S, 2016MNRAS.461.1927P, 2017ApJ...835..144G, 2019ApJ...874....8M}. Our changing look analogues are identified as variable AGN in SDSS Stripe 82, but have no variability in their (well-sampled, sufficiently high signal-to-noise) PTF light curves. These systems comprise 8/52 galaxies. All but one of the 8 are broad-line AGN based on their SDSS spectroscopy. Their stellar masses range from $9\times10^{9}-9\times10^{10}~M_{\odot}$. An example is shown in Figure~\ref{s82_ptf_cl}.

We observed three of these systems with the Echellette Spectrograph and Imager on Keck II in September 2019 and found that the broad emission had substantially faded in all three cases. Detailed analysis of these systems will be the subject of a forthcoming paper. Nevertheless, 7/41 broad line AGN ($17\%$) from \cite{2018ApJ..868..152} show substantial changes in their variability properties over time scales of $\sim10-20$ years. If all of these systems also undergo changes in their broad emission line profiles on these timescales, this result has severe consequences for the estimation of BH masses based on broad H$\alpha$ emission. A similar warning was given by \cite{2016ApJ...821...33R}. They studied the optical spectra of 102 nearby ($z<0.1$) Seyfert galaxies and found that two-thirds showed changes in either the width or flux of broad H$\beta$ on timescales of 3-9 years, with broad H$\beta$ virtually disappearing in three of them. Our NSA sample includes light curves for 8 galaxies from their sample (most of their sample is at z>0.055). They find 5 of these to show significant changes in broad H$\beta$. We find that 4/5 of those show AGN-like variability in their PTF light curves. More analysis, including light curve construction for all 102, would be needed to study any possible link between broad line appearance/disappearance and optical variability. 

Finally, for the remaining 12/52 galaxies which overlap between  \cite{2018ApJ..868..152} and this work, we cannot state definitively which of the three classes they fit in to. 

It is also interesting to consider systems where the variability turns on, i.e., galaxies which are not variable in Stripe 82 but are variable in PTF. There are \textcolor{black}{1888} non-variable galaxies from the Stripe 82 sample that are covered in our PTF sample. Of those, we find 15 to be variable in PTF, or $0.8\%$. Three are narrow-line AGN, 7 are narrow-line SF, and 5 have no BPT class due to absorption or low S/N lines. Follow-up observations of these systems is needed to explore whether this is reflected in their optical spectra.

\section{Relations with BH mass}

In this section, we explore possible relations between BH mass and variability properties. While possible relations between variability and AGN luminosity, Eddington ratio, and BH mass have been studied \citep{2007MNRAS.375..989W, 2010ApJ...721.1014M, 2012ApJ...758..104Z, 2017ApJ...834..111C}, the field has not reached a consensus. Here, we search for relations between BH mass and damped random walk parameters. A damped random walk (also known as an Ornstein-Uhlenbeck process) with noise has parameters $\mu$, $\sigma$, $\tau$, and $\omega$, where $\mu$ is the long-term mean, $\tau$ is a reversion time scale, $\sigma$ represents the instantaneous variability amplitude, and $\omega$ is the noise amplitude. These parameters can be estimated using the maximum likelihood estimator. We estimate $\mu$, $\sigma$, $\tau$, and $\omega$ using the logarithmic likelihood function given in \cite{RePEc:arx:papers:1811.09312} (see their equations 28 and 29).  

\begin{figure*}
\centering
\includegraphics[width=0.48\textwidth]{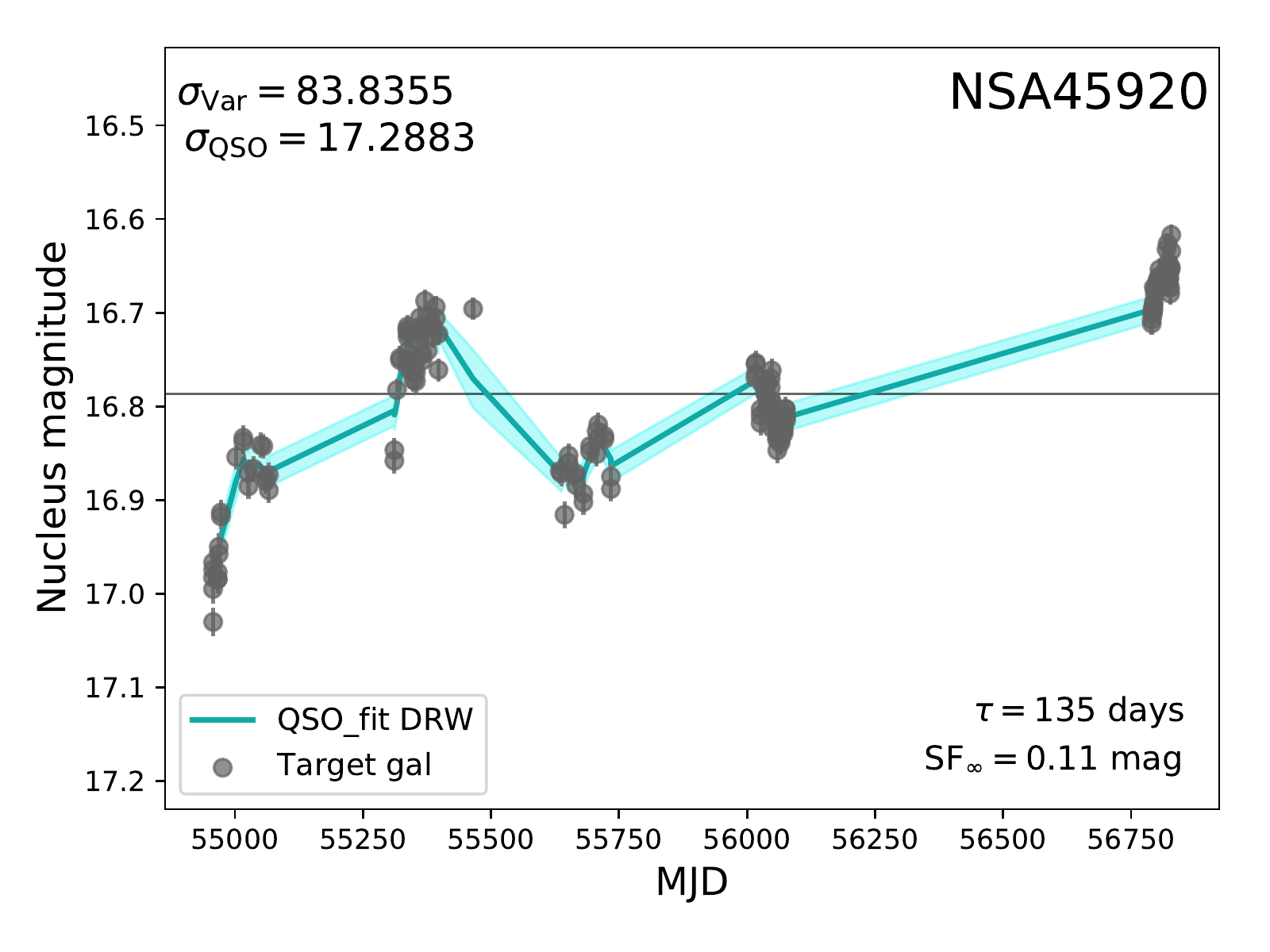}
\includegraphics[width=0.48\textwidth]{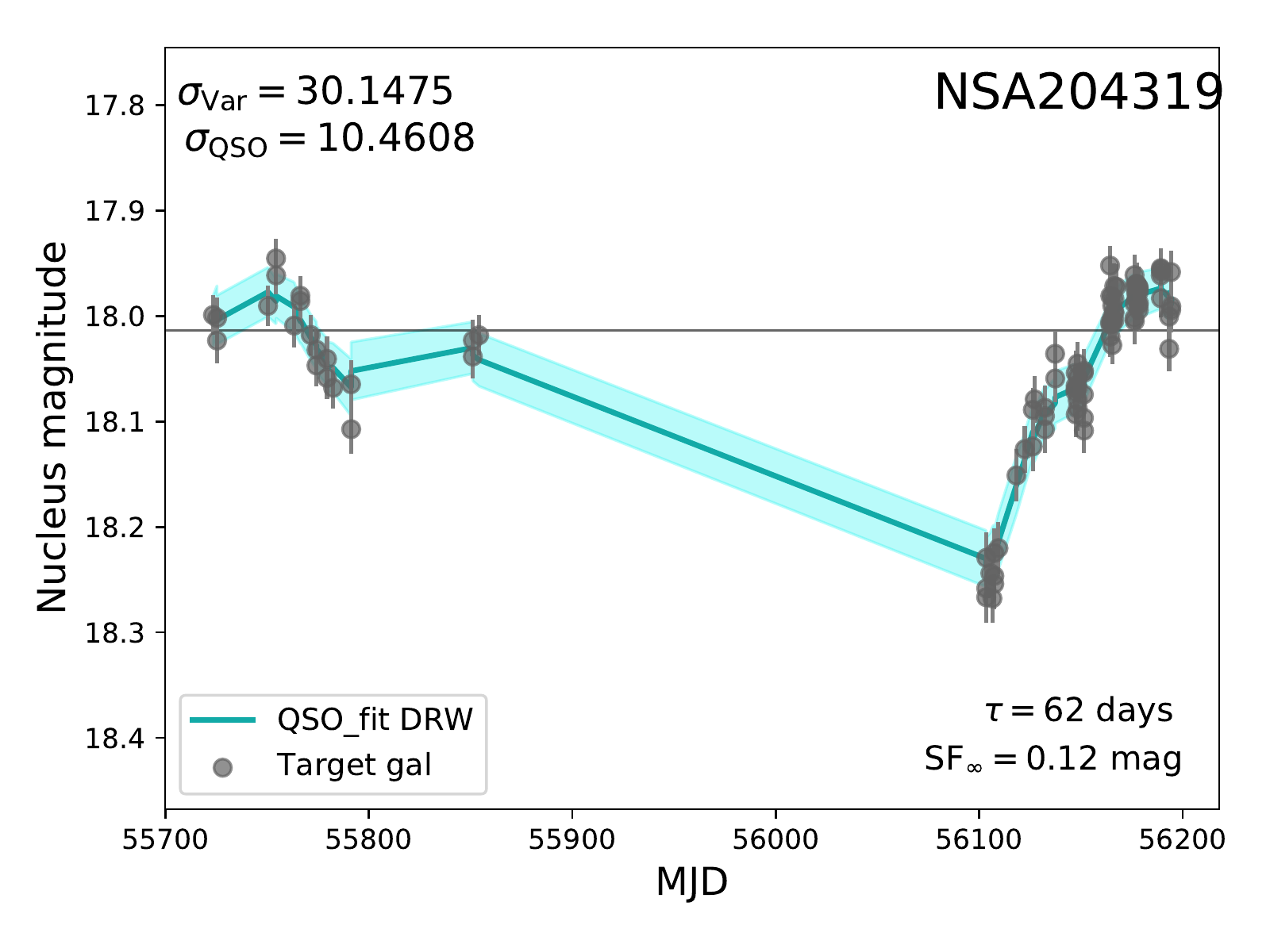}
\caption{Light curves for two broad line AGN for which we estimated the DRW parameters $\tau$ and $\rm{SF_{\infty}}$.}
\label{DRWexamp}
\end{figure*}

When fitting AGN light curves with a damped random walk, it is common to report the ``structure function at infinity" (${\rm SF_{\infty}}$); this quantity is a function of the reversion time scale $\tau$ and instantaneous variability amplitude $\sigma$ as ${\rm SF_{\infty}} = \sigma \tau^{1/2}$, and can be thought of as an asymptotic variability amplitude. Figure~\ref{DRWexamp} shows light curves of two broad line AGN and their estimated DRW parameters. 

\begin{figure*}
\centering
\includegraphics[width=0.8\textwidth]{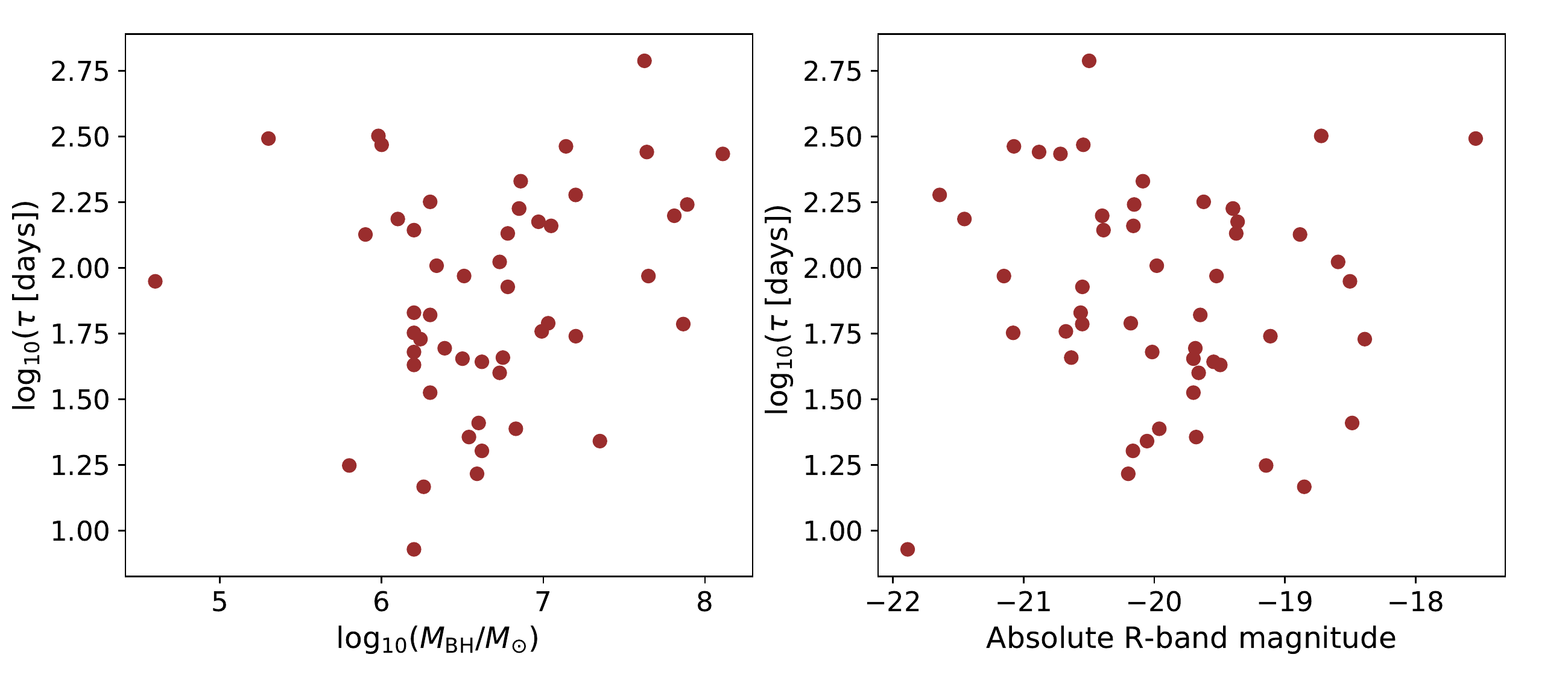}
\caption{Reversion timescale $\tau$ versus BH mass (left) and absolute PTF R-band magnitude (right). Typical uncertainties on the single epoch spectroscopic BH masses are 0.3 dex. We find no relation between the reversion timescale and BH mass or nucleus absolute magnitude.}
\label{taufig}
\end{figure*}

\begin{figure*}
\centering
\includegraphics[width=0.8\textwidth]{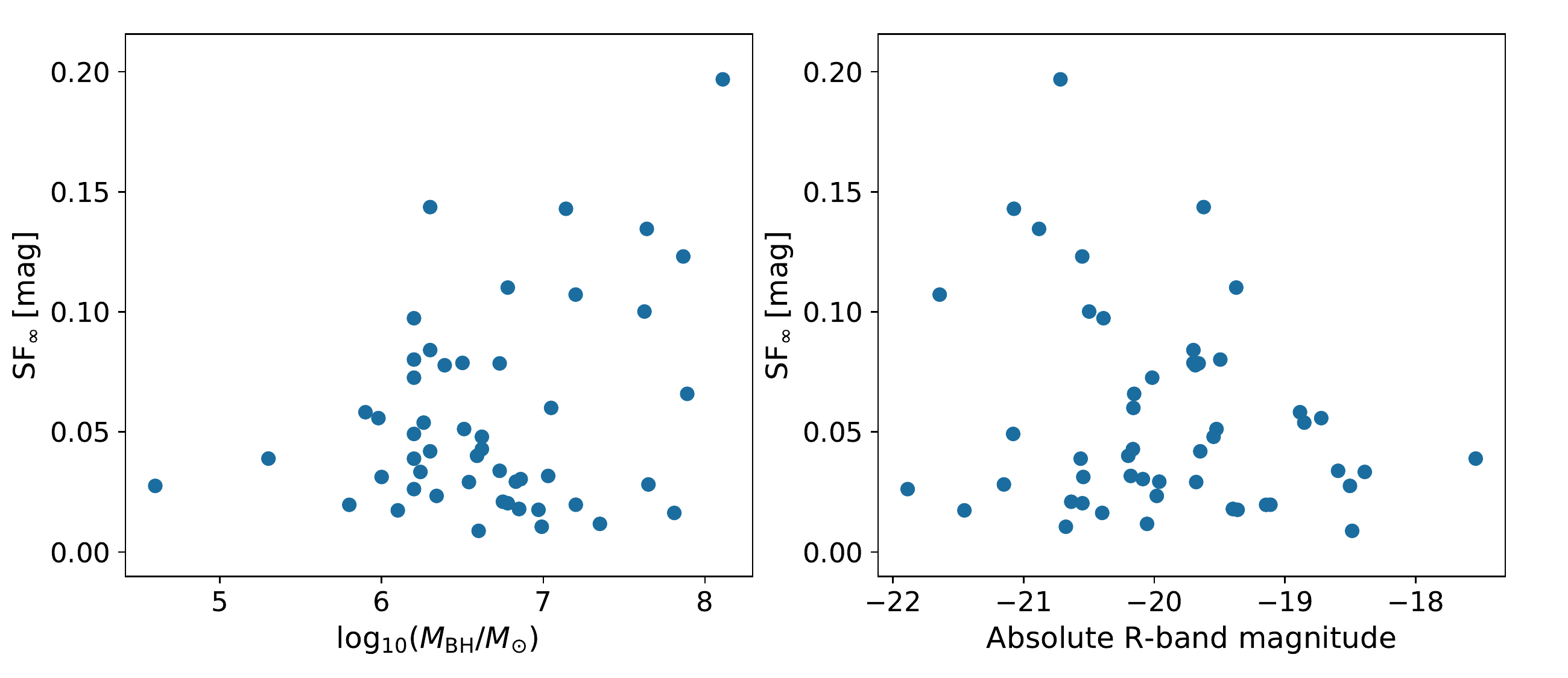}
\caption{Structure function at infinity ($\rm{SF}_{\infty}$) versus BH mass (left) and absolute PTF R-band magnitude (right). Typical uncertainties on the single epoch spectroscopic BH masses are 0.3 dex. We find no relation between $\rm{SF}_{\infty}$ and BH mass or nucleus absolute magnitude.}
\label{sffig}
\end{figure*}

We estimate $\tau$ and  ${\rm SF_{\infty}}$ for the 45 variable broad-line AGN from \cite{2007ApJ...670...92G} and \cite{2015ApJ...813...82R}, plus any of the variable AGN from our main sample which have broad H$\alpha$ emission. This gives 67 objects; $\tau$ and ${\rm SF_{\infty}}$ are well-constrained for 52 of them. We find values of $\tau$ ranging from 8 days to 600 days (median $\tau$ = 67 days) and  ${\rm SF_{\infty}}$ ranging from 0.01 to 0.2 mag (median ${\rm SF_{\infty}}$=0.04 mag).  In Figure~\ref{taufig} we show the estimated reversion timescale versus BH mass and nucleus R-band absolute magnitude. In Figure~\ref{sffig}, we plot ${\rm SF_{\infty}}$ versus black hole mass and absolute magnitude. We do not find correlations between either $\tau$ or ${\rm SF_{\infty}}$ and the BH mass and nucleus magnitude. Note that nucleus magnitude includes both the AGN and the underlying stellar population within 3$''$. 

\section{Variable AGN in low-mass galaxies}

We find 244 variability selected AGN with stellar masses below $10^{10}~M_{\odot}$. Here, we compare the population of AGN detected in low-mass galaxies via long-term variability to those detected via optical spectroscopy. We also discuss our results in the context of the low-mass end of the BH occupation fraction. 

\subsection{Comparison to other selection techniques}

\begin{figure*}
\includegraphics[width=0.5\textwidth]{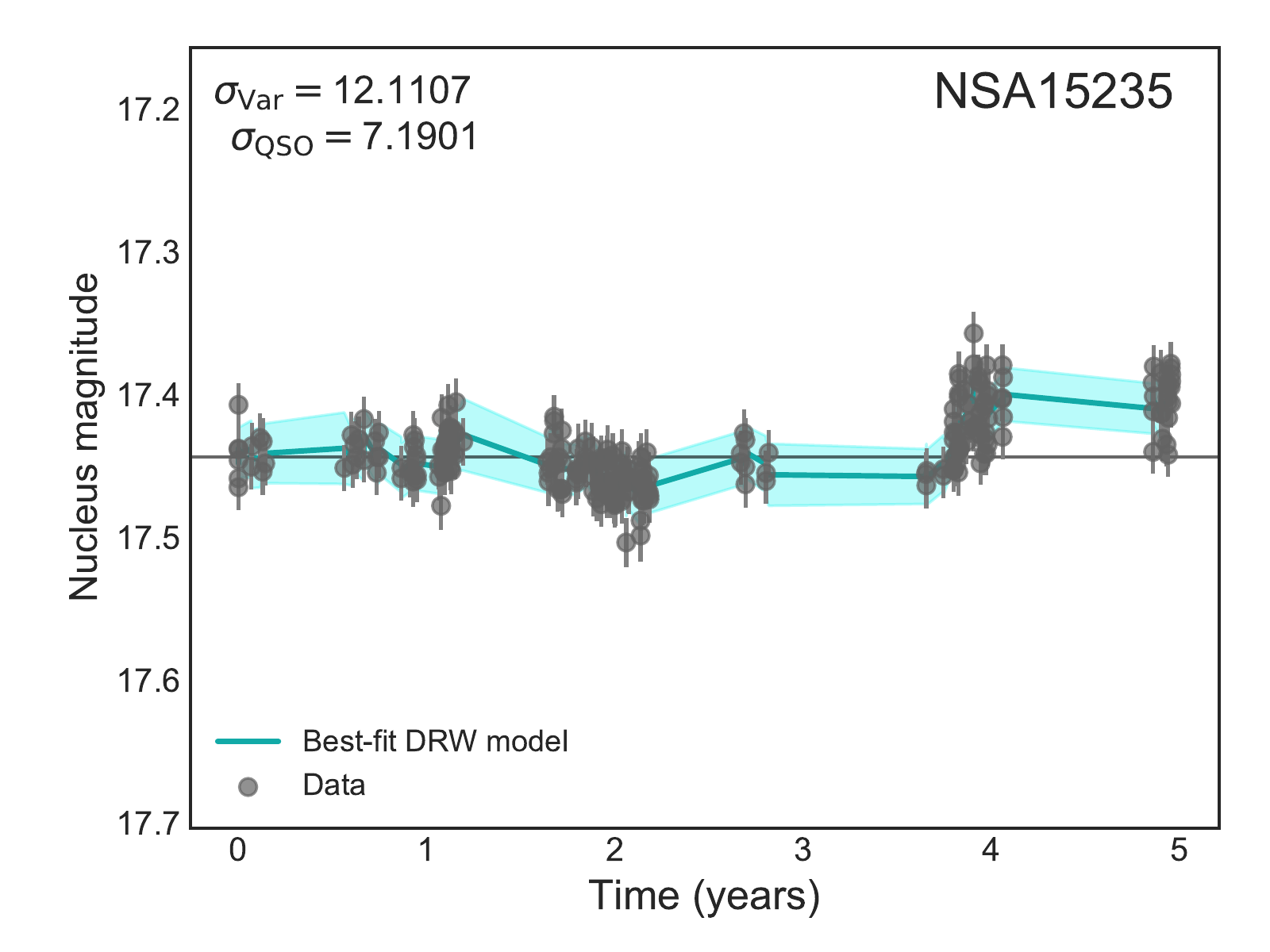}
\includegraphics[width=0.5\textwidth]{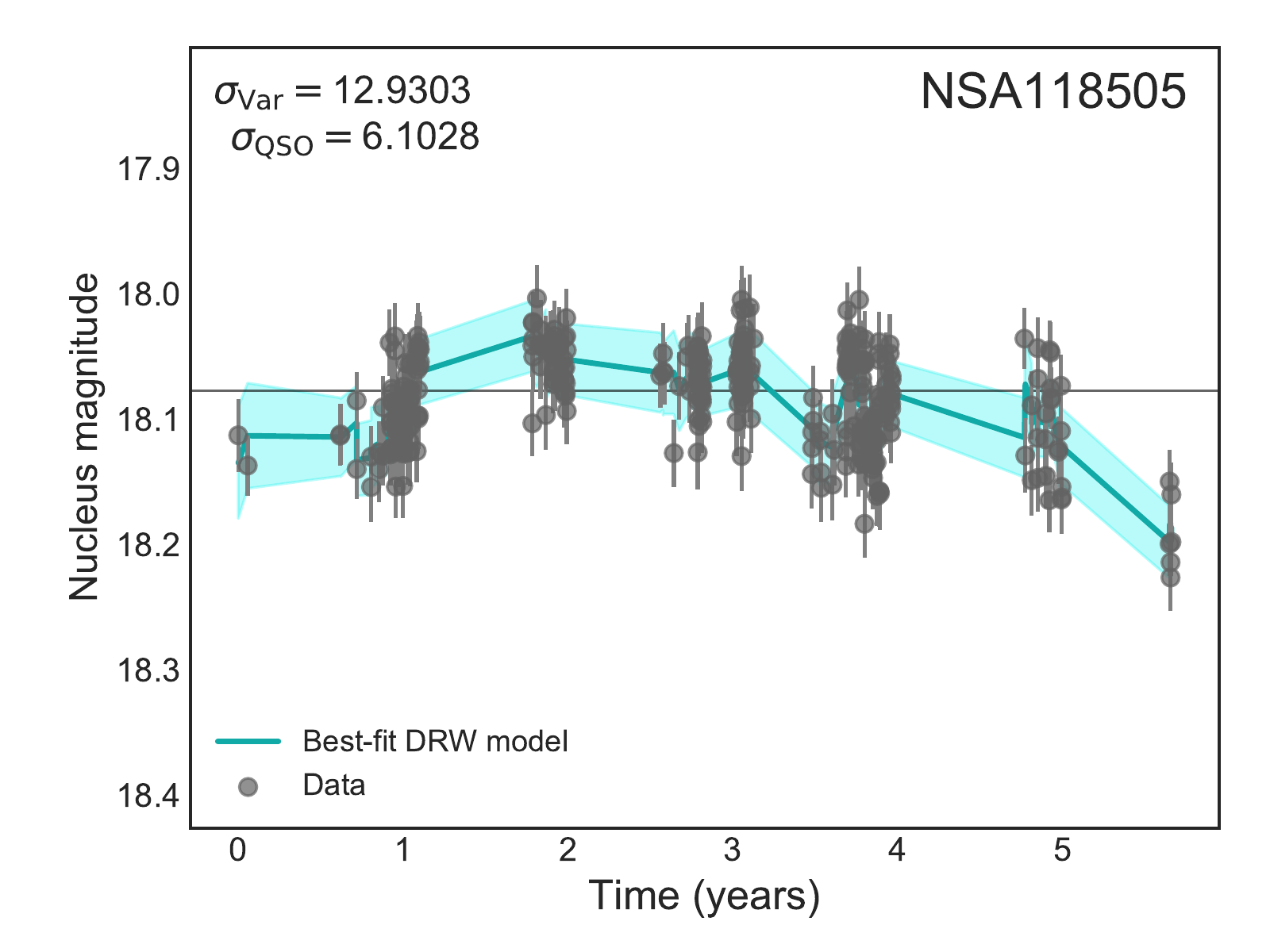}
\caption{The two galaxies from \cite{Reines:2013fj} which are found to have optical photometric variability consistent with an AGN. NSA 15235 has narrow and broad line evidence for an AGN, while NSA 118505 is a narrow-line AGN. }
\label{RGGlcs}
\end{figure*}

Most AGN in low-mass galaxies have been identified through optical spectroscopic selection techniques \citep{2004ApJ...610..722G, 2007ApJ...670...92G, Reines:2013fj, 2014AJ....148..136M}. \cite{Reines:2013fj} identified 136 dwarf galaxies (defined as galaxies with stellar masses less than the Large Magellanic Cloud, or $M_{\ast}<3\times10^{9}~M_{\odot}$) with narrow emission lines consistent with the presence of an AGN based on the BPT diagram \citep{1981PASP...93....5B}. Of the 136 galaxies with narrow-line evidence for an AGN, 10 also had broad emission lines. We are able to construct PTF light curves with more than 20 data points for 66 out of 136. We find 2/66 have AGN-like variability: NSA 15235 and NSA 118505  (RGG 32 and RGG 91, respectively, in \citealt{Reines:2013fj}). NSA 15235 has broad H$\alpha$ emission and narrow emission lines in the AGN region of the BPT diagram, while NSA 118505 has no detectable broad line in SDSS and narrow emission lines in the AGN region of the BPT diagram. The broad emission line in NSA 15235 gives a BH mass of $M_{\rm BH}\approx 1.5\times10^{5}~M_{\odot}$. Light curves for these objects are shown in Figure~\ref{RGGlcs}. 

The overall variable fraction for the \cite{Reines:2013fj} narrow-line AGN is $3^{+5}_{-2}\%$ (90\% confidence limits), consistent with the fraction of variable AGN among all narrow-line AGN/composites in our main NSA sample ($1.7\%\pm{0.2\%}$). AGN variability in dwarf galaxies with AGN-dominated narrow-emission lines is thus consistent with what is found for higher mass systems. 

We now compare the fractions of AGN in dwarf galaxies discovered via optical spectroscopy versus AGN variability. There are 44594 galaxies with stellar masses less than $3\times10^{9}~M_{\odot}$ in the NSAv0; 136 have evidence for an AGN based on the BPT diagram \citep{Reines:2013fj}. This gives an active fraction based on optical spectroscopy of $0.3\%\pm{0.05}\%$. For the same mass range ($M_{\ast}<3\times10^{9}~M_{\odot}$), we find \textcolor{black}{102} objects with AGN-like variability, out of  \textcolor{black}{18251}. This gives a variability-based active fraction of $0.56\pm{0.07}\%$. There is only a small amount of overlap between the two samples, as only 2 of our variability-selected AGN in this mass regime are also identified as AGN in \cite{Reines:2013fj}. Variability is thus identifying a complementary population of AGN in low-mass galaxies as compared to optical spectroscopic selection. 

\begin{figure}
\includegraphics[width=0.5\textwidth]{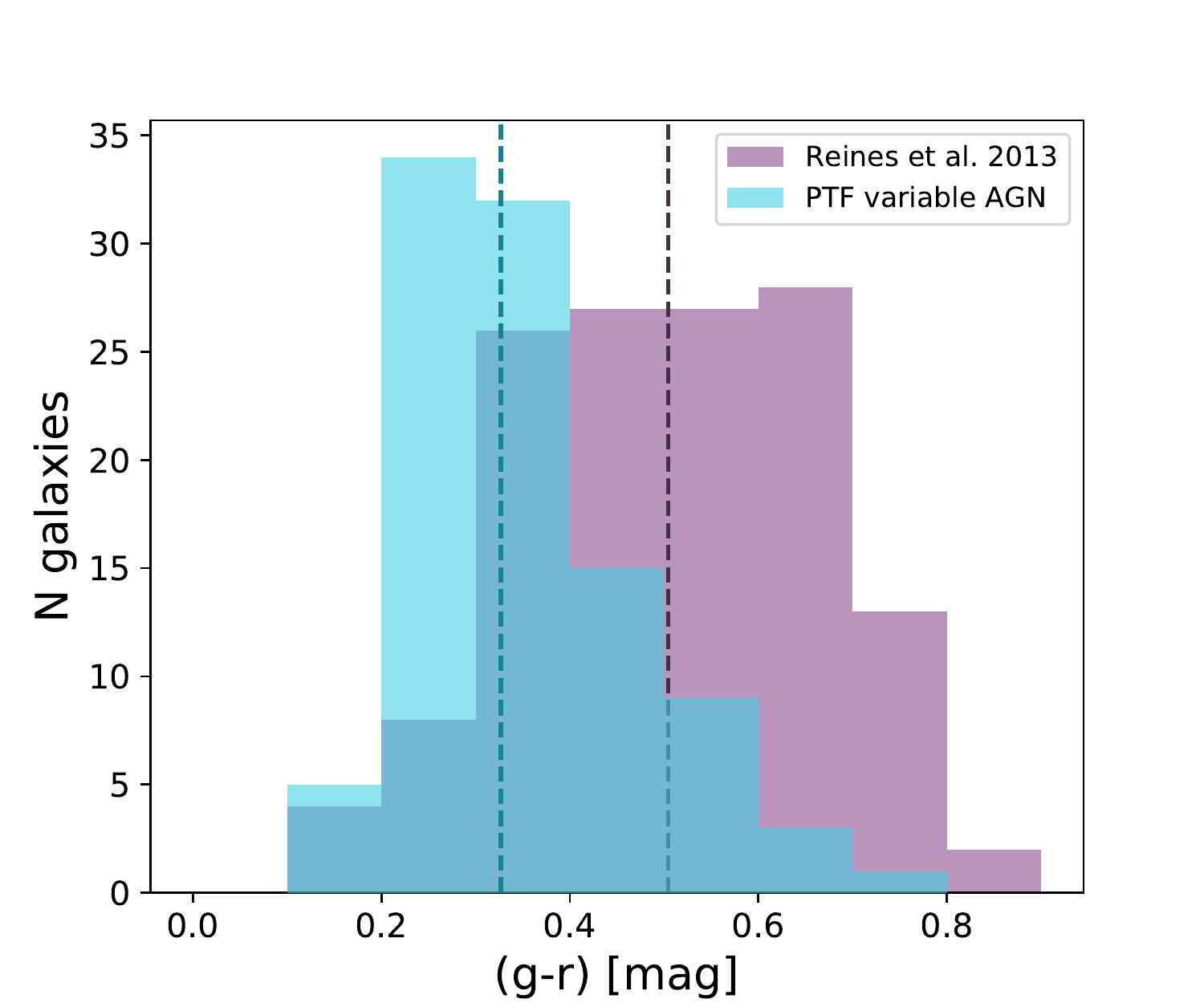}
\caption{Histogram of galaxy $g-r$ colors for optical spectroscopic-selected AGN in dwarf galaxies ($M_{\ast}<3\times10^{9}~M_{\odot}$) from \cite{Reines:2013fj} and PTF variability-selected AGNs in the same mass range. PTF variable AGN are shown in blue and \cite{Reines:2013fj} is shown in purple. The blue and purple dashed lines represent the median $g-r$ for the PTF sample and \cite{Reines:2013fj} sample, respectively. The \cite{Reines:2013fj} AGNs are, on average, in redder galaxies than the variability-selected AGNs. }
\label{RGG_comp}
\end{figure}

As discussed above, star formation dilution and/or metallicity effects could be responsible the variability-selected AGN falling primarily in the star forming region of the BPT diagram. In Figure~\ref{RGG_comp}, we compare the galaxy $g-r$ colors for variability selected AGNs to the \cite{Reines:2013fj} AGNs. The variability-selected AGNs tend to be bluer than the \cite{Reines:2013fj} AGN host galaxies (median $g-r =0.33$, compared to 0.5 for the \citealt{Reines:2013fj} AGNs), which lends support to the star formation dilution hypothesis. In some cases, isolating emission from the nucleus with higher spatial resolution spectroscopy reveals a change in the narrow line ratios towards the AGN region of the BPT diagram (e.g., \citealt{2019arXiv190201401D}). 

Another intriguing possibility is that many of the observed AGNs in dwarf galaxies come from the tidal disruption of stars. As discussed in \cite{2019MNRAS.483.1957Z}, feeding BHs in dwarf galaxies with inflowing gas streams is difficult due to stellar feedback, but the population of AGNs in dwarf galaxies could be explained by tidal disruption events (TDEs) if the occupation fraction is close to 1. Less inflowing gas in the vicinity of the BH (as compared to AGN in more massive galaxies) could explain the lack of strong broad and/or narrow AGN emission line signatures in the variability-selected AGN in low mass galaxies. Interestingly, \cite{2019MNRAS.487.4136W} studied the optical spectra of the host galaxies of $\sim25$ optical and X-ray TDEs and found that the majority of hosts were either quiescent or had narrow emission lines consistent with star formation. Spatially resolved spectroscopy of the variability-selected low-mass AGNs will reveal whether star formation dilution is indeed responsible for the star formation-dominated SDSS spectra. 

\subsection{Implications for the occupation fraction}

The occupation fraction is defined as the fraction of galaxies containing a central BH. Dynamical BH detections in nearby galaxies have revealed that the occupation fraction is $\sim100\%$ for galaxies with stellar masses greater than $10^{10}~M_{\odot}$. The occupation fraction for systems with $M_{\ast}<10^{10}~M_{\odot}$ remains relatively unconstrained, though recent estimates place it between 30-90\% \citep{2015ApJ...799...98M,2018ApJ...858..118N, 2019ApJ...872..104N}. At face value, the drop in the fraction of variable AGN towards low stellar masses suggests a lower occupation fraction for galaxies below $10^{10}~M_{\odot}$. However, as demonstrated by Figure~\ref{std_v_mag}, lower levels of fractional variability are more detectable in brighter nuclei. In order to compare high and low mass galaxies, we first must correct for differing magnitude distributions. Note that the typical baseline and number of data points for galaxies above and below $10^{10}~M_{\odot}$ are consistent with one another.

\begin{figure}
\centering
\includegraphics[width=0.5\textwidth]{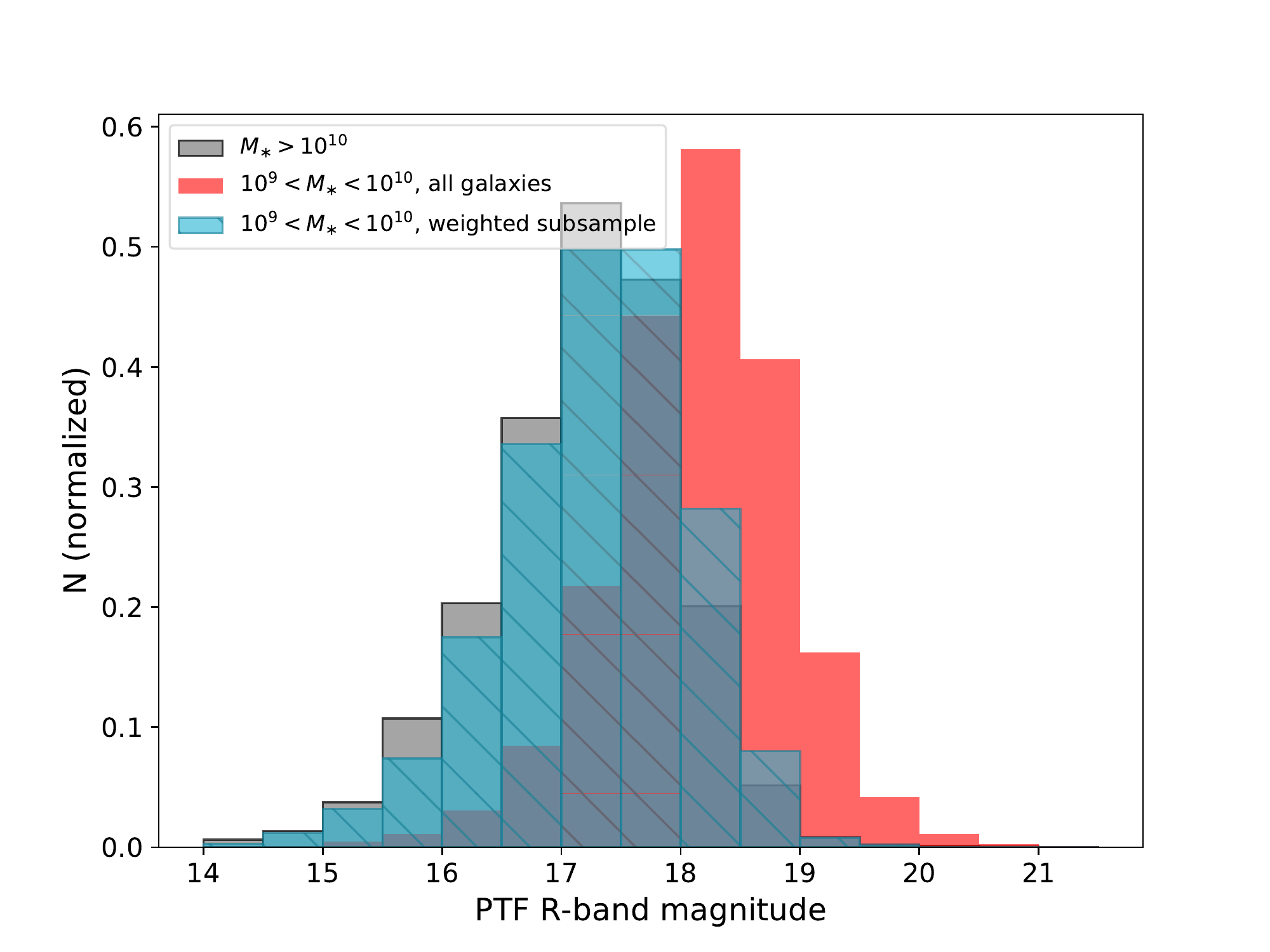}
\caption{Normalized histograms for galaxies in our sample with $M_{\ast}>10^{10}M_{\odot}$ (gray), and $10^{9}M_{\odot}<M_{\ast}<10^{10}M_{\odot}$ (red). The distribution for a low-mass magnitude-matched subsample is shown by the blue hatched histogram. We use these magnitude-matched subsamples to compare the fractions of variable AGNs in different mass regimes.  }
\label{magcomp_hist}
\end{figure}

From our main NSAv0 galaxy sample, we find that the variable AGN fraction for galaxies with $M_{\ast}>10^{10}~M_{\odot}$ is $1.5\%^{+0.17}_{-0.13}$ (90\% confidence limits).  We next consider the population between $10^{9}-10^{10}~M_{\odot}$. Since the high and low-mass samples have different apparent magnitude distributions, we create a weighting function equal to the ratio of the magnitude probability distributions for the high and low-mass samples. We then assign weights to each galaxy in the low-mass sample, and draw a random population which matches the magnitude distribution of the high-mass sample. We draw 1000 sub-samples in this manner and use the results to compute the median number of variability-selected AGN in the low-mass sample. Figure~\ref{magcomp_hist} shows an example of the magnitude distributions, demonstrating that the magnitude distribution of one instance of the low-mass subsample matches the magnitude distribution of the high-mass sample. We also repeat these step for the sample of galaxies with $M_{\ast}<10^{9}~M_{\odot}$.

We find that the magnitude-matched fraction of variable AGN for galaxies with stellar masses from $10^{9}-10^{10}~M_{\odot}$ is $1.4\%\pm0.4$ (90\% confidence limits). Within the uncertainties, this is consistent with the fraction of variable AGN in higher mass galaxies. Given that there do not appear to be differences in variability properties with BH mass, this suggests that the local occupation fraction between $10^{9}-10^{10}~M_{\odot}$ is not substantially lower than the occupation fraction above $M_{\ast}>10^{10}~M_{\odot}$. This is consistent with results based on dynamical detections of BHs in very nearby, low-mass early-type galaxies \citep{2018ApJ...858..118N, 2019ApJ...872..104N}.

For the lowest mass bin ( $M_{\ast}<10^{9}~M_{\odot}$), we find a variability fraction of $0.6\%^{+0.6}_{-0.4}$. This is inconsistent with the fraction of variable AGN in galaxies with $M_{\ast}>10^{10}~M_{\odot}$, even after accounting for different magnitude distributions. This could reflect a lower occupation fraction, but could also be due to incompleteness in the NSAv0 for galaxies with $M_{\ast}<10^{9}~M_{\odot}$, or to these galaxies hosting BHs that are below our detection thresholds. Larger and deeper samples will be needed to place meaningful constraints on the active fraction for $M_{\ast}<10^{9}~M_{\odot}$. However, there are presently no constraints on the occupation fraction in this mass regime, so identifying any galaxies with AGNs in this mass regime represents a step forward.

\section{Summary and Future Directions}

We analyze light curves of \textcolor{black}{47125} nearby (z<0.055) galaxies from the NASA-Sloan Atlas with Palomar Transient Factory coverage with the goal of identifying variable AGN in low-mass galaxies. Our sample is complete for stellar masses less than $M_{\ast}=2\times10^{10}~M_{\odot}$, but extends up to stellar masses of $10^{12}~M_{\odot}$. Our key results are summarized below.

\begin{itemize}

\item We find \textcolor{black}{424} galaxies with AGN-like variability, \textcolor{black}{244} of which have stellar masses less than $10^{10}~M_{\odot}$.

\item 75\% of low-mass galaxies identified as having AGN variability have narrow emission lines dominated by star formation, and thus would be missed in optical spectroscopic searches. This could be due to star formation dilution, low galaxy metallicity, or a difference in primary AGN fueling mechanisms for low-mass galaxies. 

\item Low-mass variability selected AGNs reside in bluer host galaxies than those selected by optical spectroscopy, with little overlap between variability and BPT-selected AGNs (Section 6.1). This suggests at least some of the variability-selected AGNs may be undetected in SDSS spectroscopy due to star formation dilution. 

\item The fraction of variable AGN is constant down to $M_{\ast}\approx10^{9}~M_{\odot}$, suggesting that the occupation fraction does not change drastically in this mass regime (Section 6.2). This is in good agreement with recent occupation fraction results from \cite{2015ApJ...799...98M} and \cite{2018ApJ...858..118N, 2019ApJ...872..104N}. 

\item Below $M_{\ast}=10^{9}~M_{\odot}$, the fraction of variable AGNs drops. It is not clear whether this is due to a change in occupation fraction or incompleteness. Larger samples and deeper repeat imaging surveys are needed to place meaningful constraints on the active fraction below stellar masses of $10^{9}~M_{\odot}$.

\item The measured AGN fraction is strongly dependent on measurement baseline. The AGN fraction for baselines less than two years is 0.25\%, compared to 1\% for baselines longer than two years.

\item We find no correlations between BH mass and the reversion timescale or variability amplitude for a sample of 51 broad-line AGN with single-epoch spectroscopic BH masses ranging from $10^{5}-10^{8}~M_{\odot}$.

\item We uncover a population of changing look-AGNs by combining Stripe 82 and PTF light curves for objects which overlap between the two samples. The combined light curves give baselines of 15-20 years. There are eight galaxies which have AGN-like variability in Stripe 82 and are quiescent in PTF, and 15 galaxies which are quiescent in Stripe 82 and variable in PTF. Spectroscopic follow-up of these changing-look systems will help determine the cause of the change in accretion properties.
\end{itemize}

The PTF survey has recently been superseded by the Zwicky Transient Facility \citep{2019PASP..131g8001G}, which has similar imaging quality and resolution, but uses a much larger CCD, facilitating faster coverage of the full sky. Future analysis will incorporate data from the ZTF survey, allowing us to study AGN variability on even longer timescales with shorter cadences. Additionally, LSST, which will image the entire visible night sky every three nights and is expected to come online in the next few years, will be an incredible resource for identifying low-mass AGN via optical variability \citep{2008SerAJ.176....1I}. 

\acknowledgements

We are grateful to the PTF collaboration and SDSS collaboration for making the data used in this paper public and easily accessible. Support for VFB was provided by the National Aeronautics and Space Administration through Einstein Postdoctoral Fellowship Award Number PF7-180161 issued by the Chandra X-ray Observatory Center, which is operated by the Smithsonian Astrophysical Observatory for and on behalf of the National Aeronautics Space Administration under contract NAS8-03060. The authors thank Claire Dickey and Michael Tremmel for useful comments which have improved this paper.

\bibliographystyle{apj}

\end{document}